%% file: flas.tex
\input harvmacMv2.tex

\input epsf.tex
\input scrload.tex

\def\Lie{{\scr L}}
\def\cL{\Lie_s}
\def\cS{{\cal S}}
\def\cL{{\cal L}}
\def\cg{\mathord{\hbox{\frakfont g}}}

\def\textbf#1{{\bf #1}}

\def\mathcal#1{{\cal #1}}
\def\frac#1#2{{#1\over #2}}
\def\proof{\par\noindent{\it Proof}.\hskip2.5pt}

\title A Lie bracket for the momentum kernel

\author
Hadleigh Frost\email{\star}{frost@maths.ox.ac.uk}$^\star$,
Carlos R. Mafra\email{\ap}{c.r.mafra@soton.ac.uk}$^\ap$ and
Lionel Mason\email{\ast}{lmason@maths.ox.ac.uk}$^\ast$

\address
$^{\star,\ast}$The Mathematical Institute, University of Oxford
Andrew Wiles Building, ROQ, Woodstock Rd, Oxford OX2 6GG, United Kingdom
\medskip
$^\ap$Mathematical Sciences and STAG Research Centre, University of Southampton,
Highfield, Southampton, SO17 1BJ, United Kingdom

\abstract
We develop new mathematical tools for the study of the double copy and colour-kinematics duality for tree-level scattering
amplitudes using the properties of Lie polynomials. We show that the $S$-map that was defined to simplify super-Yang--Mills
multiparticle superfields is in fact a new Lie bracket on the dual space of Lie polynomials. We introduce {\it Lie polynomial
currents} based on Berends-Giele recursion for biadjoint scalar tree amplitudes that take values in Lie
polynomials. Field theory amplitudes are obtained from the Lie polynomial amplitudes by numerators characterized as
homomorphisms from the free Lie algebra to kinematic data. Examples are presented for the biadjoint scalar, Yang--Mills theory
and the nonlinear sigma model. That these theories satisfy the Bern-Carrasco-Johansson amplitude relations follows from the
identities we prove for the Lie polynomial amplitudes and the existence of BCJ numerators.

A KLT map from Lie polynomials to their dual is obtained by nesting the S-map Lie bracket; the matrix elements of
this map yield a recently proposed `generalized KLT matrix', and this reduces to the usual KLT matrix when its entries are
restricted to a basis. Using this, we give an algebraic proof for the cancellation of double poles in the KLT formula for
gravity amplitudes. We finish with some remarks on numerators and colour-kinematics duality from this perspective.

\Date {December 2020}

\lref\dufufeng{
C.-H. Fu, Y.-J. Du, R.~Huang, and B.~Feng, ``Expansion of
  Einstein-Yang-Mills Amplitude'',  {\it JHEP} {\bf 09} (2017) 021,
[arXiv:1702.0815].
}

\lref\garsia{
	A.M. Garsia, ``Combinatorics of the Free Lie Algebra and the Symmetric Group'',
	In Analysis, et Cetera, edited by Paul H. Rabinowitz and E. Zehnder,
	Academic Press, (1990) 309-382
}
\lref\patras{
	F.~Patras, C.~Reutenauer, M.~Schocker, ``On the Garsia Lie Idempotent'',
   	Canad. Math. Bull. {\bf 48} (2005), 445-454
}
\lref\NLSM{
	J.J.M.~Carrasco, C.R.~Mafra and O.~Schlotterer,
	``Abelian Z-theory: NLSM amplitudes and $\alpha$'-corrections from the open string,''
	JHEP \textbf{06}, 093 (2017)
	[arXiv:1608.02569 [hep-th]].
}
\lref\kiermaier{
Gravity as the Square of Gauge Theory, talk at Amplitudes, Queen Mary, University of London
M. Kiermaier
http://www.strings.ph.qmul.ac.uk/~theory/Amplitudes2010/Talks/MK2010.pdf
}
\lref\DuNLSM{
	Y.~J.~Du and C.~H.~Fu,
	``Explicit BCJ numerators of nonlinear simga model,''
	JHEP \textbf{09}, 174 (2016)
	[arXiv:1606.05846 [hep-th]].
}
\lref\BGap{
	C.~R.~Mafra and O.~Schlotterer,
  	``Non-abelian $Z$-theory: Berends-Giele recursion for the $\alpha'$-expansion of disk integrals,''
	[arXiv:1609.07078 [hep-th]].
}
\lref\drinfeld{
	J.~Broedel, O.~Schlotterer, S.~Stieberger and T.~Terasoma,
  	``All order $\alpha^{\prime}$-expansion of superstring trees from the Drinfeld associator,''
	Phys.\ Rev.\ D {\bf 89}, no. 6, 066014 (2014).
	[arXiv:1304.7304 [hep-th]].
}
\lref\nptTreeI{
	C.~R.~Mafra, O.~Schlotterer and S.~Stieberger,
	``Complete N-Point Superstring Disk Amplitude I. Pure Spinor Computation,''
	Nucl.\ Phys.\ B {\bf 873}, 419 (2013).
	[arXiv:1106.2645 [hep-th]].
}
\lref\BGschocker{
	M. Schocker,
	``Lie elements and Knuth relations,'' Canad. J. Math. {\bf 56} (2004), 871-882.
	[math/0209327].
}
\lref\DPellis{
	F.~Cachazo, S.~He and E.Y.~Yuan,
	``Scattering of Massless Particles: Scalars, Gluons and Gravitons,''
	JHEP {\bf 1407}, 033 (2014).
	[arXiv:1309.0885 [hep-th]].
}
\lref\CBCJ{F.~Cachazo,
``Fundamental BCJ Relation in N=4 SYM From The Connected Formulation,''
[arXiv:1206.5970].}
\lref\CHYKLT{F.~Cachazo, S.~He and E.~Y.~Yuan,
``Scattering equations and Kawai-Lewellen-Tye orthogonality,''
Phys. Rev. D \textbf{90} (2014) no.6, 065001
[arXiv:1306.6575].}
\lref\Du{
Y.~J.~Du, B.~Feng and C.~H.~Fu,
``BCJ Relation of Color Scalar Theory and KLT Relation of Gauge Theory,''
JHEP \textbf{08} (2011), 129
[arXiv:1105.3503 [hep-th]].
}
\lref\michos{
	Michos, Ioannis C. 
       ``On twin and anti-twin words in the support of the free Lie algebra.''
       Journal of Algebraic Combinatorics {\bf 36.3} (2012), 355-388.
       [arXiv:1011.1528 [math.CO]].
}
\lref\PScomb{
	C.R.~Mafra,
	``Planar binary trees in scattering amplitudes.''
	Algebraic Combinatorics, Resurgence, Moulds and Applications (CARMA) (2020): 349-365.
	[arXiv:2011.14413 [math.CO]].
}
\lref\Frost{
H.~Frost and L.~Mason,
``Lie Polynomials and a Twistorial Correspondence for Amplitudes,''
[arXiv:1912.04198 [hep-th]].
}
\lref\Mizcop{
S.~Mizera,
``Kinematic Jacobi Identity is a Residue Theorem: Geometry of Color-Kinematics Duality for Gauge and Gravity Amplitudes,''
Phys. Rev. Lett. \textbf{124} (2020) no.14, 141601
[arXiv:1912.03397 [hep-th]].
}
\lref\EST{A.~Edison, S.~He, O.~Schlotterer and F.~Teng,
``One-loop Correlators and BCJ Numerators from Forward Limits,''
JHEP \textbf{09} (2020), 079
[arXiv:2005.03639 [hep-th]].
}
\lref\ET{A.~Edison and F.~Teng,
``Efficient Calculation of Crossing Symmetric BCJ Tree Numerators,''
[arXiv:2005.03638 ].
}
\lref\MafraKJ{
	C.R.~Mafra, O.~Schlotterer and S.~Stieberger,
  	``Explicit BCJ Numerators from Pure Spinors,''
	JHEP {\bf 1107}, 092 (2011).
	[arXiv:1104.5224 [hep-th]].
}
\lref\genredef{
	E.~Bridges and C.R.~Mafra,
	``Algorithmic construction of SYM multiparticle superfields in the BCJ gauge,''
	JHEP \textbf{10}, 022 (2019)
	[arXiv:1906.12252 [hep-th]].
}
\lref\FTlimit{
	C.R.~Mafra,
  	``Berends-Giele recursion for double-color-ordered amplitudes,''
	JHEP {\bf 1607}, 080 (2016).
	[arXiv:1603.09731 [hep-th]].
}
\lref\Reutenauer{
	C.~Reutenauer,
	``Free Lie Algebras,''
	London Mathematical Society Monographs, 1993
}
\lref\Ree{
	R.~Ree, ``Lie elements and an algebra associated with shuffles'',
	Ann.Math. {\bf 62}, No.2 (1958), 210--220.
}
\lref\lothaire{
	Lothaire, M., ``Combinatorics on Words'',
	(Cambridge Mathematical Library), Cambridge University Press (1997).
}
\lref\EOMbbs{
	C.R.~Mafra and O.~Schlotterer,
  	``Multiparticle SYM equations of motion and pure spinor BRST blocks,''
	JHEP {\bf 1407}, 153 (2014).
	[arXiv:1404.4986 [hep-th]].
}
\lref\bandieraPC{
	R.~Bandiera, private communication.
}
\lref\selivanov{
	K.G.~Selivanov,
  	``On tree form-factors in (supersymmetric) Yang-Mills theory,''
	Commun.\ Math.\ Phys.\  {\bf 208}, 671 (2000).
	[hep-th/9809046].
	\semi
	K.G.~Selivanov,
  	``Postclassicism in tree amplitudes,''
	[hep-th/9905128].
}
\lref\Gauge{
	S.~Lee, C.R.~Mafra and O.~Schlotterer,
  	``Non-linear gauge transformations in $D=10$ SYM theory and the BCJ duality,''
	JHEP {\bf 1603}, 090 (2016).
	[arXiv:1510.08843 [hep-th]].
}
\lref\BCJ{
	Z.~Bern, J.J.M.~Carrasco and H.~Johansson,
	``New Relations for Gauge-Theory Amplitudes,''
	Phys.\ Rev.\ D {\bf 78}, 085011 (2008).
	[arXiv:0805.3993 [hep-ph]].
}
\lref\BCJdc{
Z.~Bern, J.~J.~M.~Carrasco and H.~Johansson,
``Perturbative Quantum Gravity as a Double Copy of Gauge Theory,''
Phys. Rev. Lett. \textbf{105} (2010), 061602
[arXiv:1004.0476 [hep-th]].
}
\lref\BGpaper{
	F.A.~Berends, W.T.~Giele,
  	``Recursive Calculations for Processes with n Gluons,''
	Nucl.\ Phys.\  {\bf B306}, 759 (1988).
}
\lref\BGsym{
	F.A.~Berends and W.T.~Giele,
	``Multiple Soft Gluon Radiation in Parton Processes,''
	Nucl.\ Phys.\ B {\bf 313}, 595 (1989).
}
\lref\BGBCJ{
	C.R.~Mafra and O.~Schlotterer,
  	``Berends-Giele recursions and the BCJ duality in superspace and components,''
	JHEP {\bf 1603}, 097 (2016).
	[arXiv:1510.08846 [hep-th]].
}
\lref\nptMethod{
	C.R.~Mafra, O.~Schlotterer, S.~Stieberger and D.~Tsimpis,
	``A recursive method for SYM n-point tree amplitudes,''
	Phys.\ Rev.\ D {\bf 83}, 126012 (2011).
	[arXiv:1012.3981 [hep-th]].
}
\lref\NovelliThibonToumazet{
	Novelli JC, Thibon JY, Toumazet F.
	``A noncommutative cycle index and new bases of quasi-symmetric functions and noncommutative
	symmetric functions.'' Annals of Combinatorics. 2020 Jul 31:1-20.
}
\lref\KKref{
	R.~Kleiss and H.~Kuijf,
	``Multi - Gluon Cross-sections and Five Jet Production at Hadron Colliders,''
	Nucl.\ Phys.\ B {\bf 312}, 616 (1989)..
}
\lref\KKproof{
	V.~Del Duca, L.J.~Dixon and F.~Maltoni,
	``New color decompositions for gauge amplitudes at tree and loop level,''
	Nucl.\ Phys.\ B {\bf 571}, 51 (2000).
	[hep-ph/9910563].
}
\lref\oldMomKer{
	Z.~Bern, L.~J.~Dixon, M.~Perelstein and J.~S.~Rozowsky,
	``Multileg one loop gravity amplitudes from gauge theory,''
	Nucl.\ Phys.\ B {\bf 546}, 423 (1999).
	[hep-th/9811140].
}
\lref\KLTmatrixrefI{
	N.~E.~J.~Bjerrum-Bohr, P.~H.~Damgaard, B.~Feng and T.~Sondergaard,
	``Gravity and Yang-Mills Amplitude Relations,''
	Phys. Rev. D \textbf{82}, 107702 (2010)
	[arXiv:1005.4367 [hep-th]].
	}
\lref\KLTmatrixrefII{
	N.~E.~J.~Bjerrum-Bohr, P.~H.~Damgaard, B.~Feng and T.~Sondergaard,
	``Proof of Gravity and Yang-Mills Amplitude Relations,''
	JHEP \textbf{09}, 067 (2010)
	[arXiv:1007.3111 [hep-th]].
}
\lref\MomKer{
	N.~E.~J.~Bjerrum-Bohr, P.~H.~Damgaard, T.~Sondergaard and P.~Vanhove,
	``The Momentum Kernel of Gauge and Gravity Theories,''
	JHEP {\bf 1101}, 001 (2011).
	[arXiv:1010.3933 [hep-th]].
}
\lref\Polylogs{
	J.~Broedel, O.~Schlotterer and S.~Stieberger,
	``Polylogarithms, Multiple Zeta Values and Superstring Amplitudes,''
	Fortsch.\ Phys.\  {\bf 61}, 812 (2013).
	[arXiv:1304.7267 [hep-th]].
}
\lref\OSpriv{O.~Schlotterer, private communication.}

\lref\BGmizera{
S.~Mizera and B.~Skrzypek,
``Perturbiner Methods for Effective Field Theories and the Double Copy,''
JHEP \textbf{10}, 018 (2018)
[arXiv:1809.02096 [hep-th]].
}
\lref\mizera{
S.~Mizera,
``Scattering Amplitudes from Intersection Theory,''
Phys. Rev. Lett. \textbf{120}, no.14, 141602 (2018)
[arXiv:1711.00469 [hep-th]].
\semi
S.~Mizera,
``Combinatorics and Topology of Kawai-Lewellen-Tye Relations,''
JHEP \textbf{08}, 097 (2017)
[arXiv:1706.08527 [hep-th]].
}
\lref\mizKLT{
S.~Mizera,
``Inverse of the String Theory KLT Kernel,''
JHEP \textbf{06}, 084 (2017)
[arXiv:1610.04230].
}
\lref\psf{
 	N.~Berkovits,
	``Super-Poincare covariant quantization of the superstring,''
	JHEP {\bf 0004}, 018 (2000)
	[arXiv:hep-th/0001035].
	\semi
	N.~Berkovits,
  	``ICTP lectures on covariant quantization of the superstring,''
	ICTP Lect.\ Notes Ser.\  {\bf 13}, 57 (2003).
	[hep-th/0209059].
}
\lref\fundamentalBCJ{
	B.~Feng, R.~Huang and Y.~Jia,
	``Gauge Amplitude Identities by On-shell Recursion Relation in S-matrix Program,''
	Phys. Lett. B \textbf{695}, 350-353 (2011)
	[arXiv:1004.3417 [hep-th]].
}
\lref\FORM{
	J.A.M.~Vermaseren,
  	``New features of FORM,''
  	math-ph/0010025.
	M.~Tentyukov and J.A.M.~Vermaseren,
  	``The multithreaded version of FORM,''
  	hep-ph/0702279.
}
\lref\monodVanhove{
	N.E.J.~Bjerrum-Bohr, P.H.~Damgaard, P.~Vanhove,
  	``Minimal Basis for Gauge Theory Amplitudes,''
	Phys.\ Rev.\ Lett.\  {\bf 103}, 161602 (2009).
	[arXiv:0907.1425 [hep-th]].
}
\lref\monodStie{
	S.~Stieberger,
  	``Open \& Closed vs. Pure Open String Disk Amplitudes,''
	[arXiv:0907.2211].
}
\lref\dushuffle{
	Y.~X.~Chen, Y.~J.~Du and B.~Feng,
	``A Proof of the Explicit Minimal-basis Expansion of Tree Amplitudes in Gauge Field Theory,''
	JHEP \textbf{02}, 112 (2011)
	[arXiv:1101.0009 [hep-th]].
}
\lref\JJreview{
	Z.~Bern, J.~J.~Carrasco, M.~Chiodaroli, H.~Johansson and R.~Roiban,
	``The Duality Between Color and Kinematics and its Applications,''
	[arXiv:1909.01358 [hep-th]].
}
\lref\KLT{
	H.~Kawai, D.~C.~Lewellen and S.~H.~H.~Tye,
	``A Relation Between Tree Amplitudes of Closed and Open Strings,''
	Nucl.\ Phys.\ B {\bf 269}, 1 (1986).
}
\lref\sigmaDu{
	G.~Chen and Y.~J.~Du,
	``Amplitude Relations in Non-linear Sigma Model,''
	JHEP \textbf{01}, 061 (2014)
	[arXiv:1311.1133 [hep-th]].
}
\lref\liecoalg{
	Michaelis, W., 1980. ``Lie coalgebras''. Advances in mathematics, 38(1), pp.1-54.
}
\lref\ABHY{
N.~Arkani-Hamed, Y.~Bai, S.~He and G.~Yan,
``Scattering Forms and the Positive Geometry of Kinematics, Color and the Worldsheet,''
JHEP \textbf{05} (2018), 096
[arXiv:1711.09102].
}
\lref\AS{
L.~Mason and D.~Skinner,
``Ambitwistor strings and the scattering equations,''
JHEP \textbf{07} (2014), 048
doi:10.1007/JHEP07(2014)048
[arXiv:1311.2564 [hep-th]].
}

\lref\FrostDPhil{H.~Frost, ``Universal aspects of perturbative gauge theory amplitudes,'' Oxford DPhil thesis, 2020.
}

\lref\CheungRW{Cheung, C., Remmen, G.N., Shen, C. et al., ``Pions as gluons in higher dimensions,''
J. High Energ. Phys. 2018, 129 (2018). https://doi.org/10.1007/JHEP04(2018)129.
}

\lref\gangchen{
G.~Chen, S.~Li and H.~Liu,
``Off-shell BCJ Relation in Nonlinear Sigma Model,''
[arXiv:1609.01832].
}
\lref\BGFthree{
	L.M.~Garozzo, L.~Queimada and O.~Schlotterer,
  	``Berends-Giele currents in Bern-Carrasco-Johansson gauge for $F^3$- and $F^4$-deformed Yang-Mills amplitudes,''
	JHEP {\bf 1902}, 078 (2019).
	[arXiv:1809.08103 [hep-th]].
}
\lref\BGTrnka{
	K.~Kampf, J.~Novotny and J.~Trnka,
  	``Recursion relations for tree-level amplitudes in the $SU(N)$ nonlinear sigma model,''
	Phys.\ Rev.\ D {\bf 87}, no. 8, 081701 (2013).
	[arXiv:1212.5224].
	K.~Kampf, J.~Novotny and J.~Trnka,
	``Tree-level Amplitudes in the Nonlinear Sigma Model,''
	JHEP {\bf 1305}, 032 (2013).
	[arXiv:1304.3048 [hep-th]].
}
\lref\andre{
	A.~Kaderli,
	``A note on the Drinfeld associator for genus-zero superstring amplitudes in twisted de Rham theory,''
	J. Phys. A \textbf{53}, no.41, 415401 (2020)
	[arXiv:1912.09406 [hep-th]].
}

\listtoc
\writetoc
\filbreak

\newsec Introduction

This paper develops Lie polynomials \Reutenauer\ and the combinatorics of words \lothaire\ as basic tools for the study of
tree-level scattering amplitudes in QFT and string theory.
We give streamlined self-contained proofs of the basic results that underpin coloured
amplitudes and the double copy at tree-level using simply the elementary properties of Lie
polynomials. We will see that cleanest description is in
terms of the Berends-Giele currents of biadjoint scalar theory, with values in Lie polynomials.

The free Lie algebra, $\Lie$, is the space of linear combinations of `Lie monomials', which are nested complete formal
commutators of `letters'; our letters will be taken to be the natural numbers, $1,2,3,\ldots$. $\Lie$ is a subspace of the space
of linear combinations of `words' formed from the natural numbers, with all letters distinct. There is a natural map from Lie
monomials $\Gamma\in \Lie$ to the colour structures that appear in gauge theories, for any choice of gauge Lie algebra, and for
any (single trace) gauge theory Lagrangian. There is also a correspondence between Lie monomials $\Gamma\in \Lie$, thought of as
complete commutators of a word, and trivalent trees with a given root.

The double copy starts by expressing Yang-Mills tree amplitudes in the form \BCJ\
\eqn\BCJamp{
A= \sum_\Gamma \frac{N_\Gamma c_\Gamma}{s_\Gamma}\, .
}
Here  $\Gamma$ denotes trivalent graphs,  $c_\Gamma$ denotes the corresponding colour factor\foot{formed from the graph by
introducing the Lie algebra's  structure constants at each vertex,  Kronecker-deltas along each internal propagator and the
external colours at the leaves.},  and $s_\Gamma$ denotes the product of denominator propagator factors associated to the graph.
The numerators $N_\Gamma$ are functions of momenta and multilinear in the gluon polarization data.  These are said to be `BCJ
numerators' if they satisfy colour kinematics duality: if $\Gamma+\Gamma'+\Gamma''= 0$, then
$N_{\Gamma}+N_{\Gamma'}+N_{\Gamma''}=0$. In other words, the $N_\Gamma$ are `BCJ numerators' if $\Gamma\mapsto N_\Gamma$ is a
homomorphism from $\Lie$ to the kinematic data.  Such numerators exist for Yang-Mills, and the key example of the double copy  is that replacing $c_\Gamma$ in \BCJamp\ by another copy of $N_\Gamma$ yields gravity amplitudes \BCJdc. BCJ numerators are known for many coloured theories and can be used to obtain the tree amplitudes of any theory known to participate in the double copy. This includes gauge and gravity theories and their relatives, such as brane theories, with and without supersymmetry, see  \JJreview\ for an up-to-date review of progress and references to the literature.  The most basic example is to replace $N_\Gamma$ in \BCJamp\ by $c_\Gamma$: this gives the amplitudes of the biadjoint scalar theory.  This theory forms the backbone of the double copy.

Lie polynomials are ubiquitous in the auxiliary structures that are used to study amplitudes, and with hindsight can be seen in
the multiparticle vertex operators in conventional string theory \EOMbbs, in the geometry of the space of Mandelstam variables
\ABHY, and in the CHY formulae and ambitwistor strings \refs{\DPellis,\AS} via the structure of the moduli space of $n$-points
on the Riemann sphere \refs{\Frost,\Mizcop}. However, our aim here is to prove basic results directly using only the Lie
polynomial structure.

The following is a
detailed summary of results.

\newsubsec\bgmethodintro Berends-Giele recursion and planar binary trees

In \S\defs\ we review the properties of the vector space of Lie polynomials, $\Lie$. Important for the applications in this
paper is the dual vector space, $\Lie^*$. Elements of $\Lie^*$ can be expressed as words $P$ modulo proper shuffles, i.e.\ the
sum $R\shuffle S$ over ordered permutations with $R,S\neq \emptyset$.

In \S\BGsec\ we introduce perturbative $\Lie$ valued fields for biadjoint scalar theory and use them to define corresponding
Lie-polynomial Berends-Giele currents.\foot{The more familiar Yang-Mills case \BGpaper\ is treated similarly in appendix \BGreview\ where the
analogous objects are $\Lie$-valued Berends Giele currents.} The Lie-polynomial currents $b(P)$ satisfy a recursion \PScomb
\eqn\introbP{
b(P)=\frac1{s_P} \sum_{XY=P} [b(X),b(Y)]\, ,
}
where $s_P$ is the Mandelstam denoting the square of the off-shell momentum of the particles in the set $\{P\}$ and the commutator is that in $\Lie$.
This defines a map
$b:\Lie^*\rightarrow \Lie_\cS$ where $\Lie^*$ encodes the colour ordering in the form of a words $P$ up to proper shuffles, and
$\Lie_\cS$ is the space of Lie polynomials with coefficients in Mandelstam variables.

The currents $b(P)$ give rise to {\it Lie polynomial amplitudes} obtained by removing
the off-shell external propagator to give single colour ordered partial amplitudes $m(Pn)$
valued in Lie polynomials:
\eqn\intrompn{
m(Pn):=\lim_{s_P\rightarrow 0} s_P b(P)\in \Lie.
}
The pairing of \intrompn\ with an ordering gives the double colour ordered partial amplitudes of the biadjoint
theory $m(Pn,Qn):=(Q,m(Pn))$ \FTlimit.

We will interpret BCJ numerators $N_\Gamma$
as homomorphisms from the free Lie algebra to appropriate functions of their kinematic data,
as in \Frost. The existence of such homomorphisms is not in principle guaranteed from this
construction, but we will explicitly write down examples for NLSM and SYM theories in \S\Atheorysec.
All identities such as the Kleiss-Kuijf (KK) relations and the Bern Carrasco
Johansson (BCJ) relations that are obeyed by $b$ are inherited by tree-level scattering amplitudes
obtained from the homomorphism.

\newsubsec\BCJintro BCJ amplitude relations and a Lie bracket

It was argued in \refs{\EOMbbs,\BGBCJ} that BCJ amplitude relations could be expressed
using the $S$-map defined in \EOMbbs. We will show that
the $S$-map corresponds to a Lie bracket $\{\, ,\}$ in the dual space of Lie polynomials and that
the BCJ amplitude relations \BCJ\ follow from the identity
\eqn\bcurlyint{
b(\{P,Q\})= [b(P),b(Q)]
}
which generalizes the off-shell BCJ  relations of \Du.
Thus $b$ maps the $\{,\}$-bracket to the standard Lie bracket.  Since $b(P)$ is invertible as a map $\Lie^*\rightarrow \Lie$, this shows
that $\{,\}$ is the pullback of $[,]$ from $\Lie_s$ to $\Lie^*_s$ using $b(P)$; in particular $\{,\}$ is a Lie bracket on
$\Lie^*$.

For theories obtained from a homomorphism acting on the Lie polynomial
amplitude \intrompn, the BCJ relation for amplitudes follow from $m(\{P,Q\},n)=0$ in the limit as
$s_{PQ}\rightarrow 0$ as $b(\{P,Q\})$ no longer has a pole in $1/s_{PQ}$ due to \bcurlyint.

\newsubsec\genKLTintro The KLT inner product and its generalized matrix

The Kawai--Lewellen-Tye (KLT) matrix \refs{\oldMomKer,\KLTmatrixrefI,\MomKer} is the inner product that is required to turn two sets
of colour ordered Yang-Mills amplitudes into a gravity amplitude \KLT.  We show that its origins can be traced to the
$\{,\}$--bracket.  The fact that $\{,\}$ forms a Lie bracket means that any Lie monomial $\Gamma$ can be rebuilt out of $\{,\}$
in $\Lie^*$ to form $\{\Gamma\}\in \Lie^*$.  This gives our most abstract definition of the KLT kernel as the map
$\Lie\rightarrow \Lie^*$ by $\Gamma\rightarrow \{\Gamma\}$; this determines the symmetric bilinear form on $\Lie$ by the pairing
$S(\Gamma_1,\Gamma_2):=(\{\Gamma_1\}, \Gamma_2)$.  We show that the standard KLT matrix is most naturally understood as the
components of this $S$ in a special `Lyndon' basis; the generalized KLT matrix of \PScomb\ is
\eqn\introSPQ{
S(P|Q)= (\ell\{P\},\ell[Q])\,,
}
where $\ell$ denotes the complete left bracketings:
\eqn\introellc{
\ell\{123\ldots n\}:=\{\ldots \{1,2\}\ldots n\}\, , \quad \ell[123\ldots n]:=[\ldots [1,2]\ldots n]\, .
}
This generalized KLT matrix is symmetric and reduces to the standard KLT matrix in a Lyndon word basis of $\Lie^*$ that singles
out a special single particle.

Cachazo, He and Yuan \DPellis\  emphasize that biadjoint scalar
amplitudes are in some sense the inverse to  the  KLT matrix, see \mizera. Using the Berends-Giele amplitude formula
for the biadjoint amplitudes of \FTlimit, this statement translates to the algebraic relation $\delta_{P,Q} = \sum_R S(P|R)_i
b(iR|iQ)$. Using that the KLT matrix corresponds to a particular basis choice of
a more fundamental KLT map, this identity will be proven in section~\genKLTsec.

\newsubsec\discussion  Numerators and cobrackets

In \S\seccontact\ we show that the ``contact term map'' defined in \genredef\ is the Lie co-bracket dual to the $S$-bracket;  it gives rise to
a {\it Lie co-algebra} structure on $\Lie^*_\cS$.
As discussed in \genredef, the contact-term map
encodes the BRST variations of local multiparticle superfield numerators  satisfying generalized
Jacobi identities \EOMbbs\ in the pure spinor formalism of the superstring \psf.
These BRST variations play a central role in all recent developments in the explicit
calculation of superstring amplitudes, from tree-level to $3$-loops.

In \S\Atheorysec\ we discuss BCJ numerators from the perspectives of this paper. We discuss the distinctions in gauge freedoms
between on-shell and Berends-Giele numerators; Berends-Giele off-shell extensions are not unique and depend on a choice of gauge
and field redefinitions.  Given such a choice, BCJ numerators are unique, but their locality will in general depend on the gauge
choices etc.. We give a brief discussion of known numerators; the biadjoint scalar case is trivial being realized by the pairing
$(P,\Gamma)$ of a Lie polynomial with a word $P$.  We give a conjecture for the non-linear sigma model (NLSM) that has now been
proved elsewhere \FrostDPhil, and briefly describe how super--Yang--Mills (sYM), Z-theory and finally the open superstring
including $\ap$ corrections fits into this framework. We end with a brief discussion of colour/kinematics duality within the
framework of the Lie brackets discussed in the text.


\newnewsec\defs Review of Lie polynomials, combinatorics on words and colour factors

Let $W(A)$ be the vector space of linear combinations of words with non-repeated 
letters in the alphabet $A\subset {\Bbb N}$.\foot{The definitions in this review section
can be stated in the free associative algebra, in which the words may contain repetition of letters. However,
in the context of scattering amplitudes we only use  words that are permutations of a subset of ${\Bbb N}$ with 
no repeated letters.} The multilinear free Lie algebra  $\Lie(A) \subset W(A)$ is  the subspace of $W(A)$
linearly generated by the Lie monomials, $\Gamma$, that are defined to be
complete bracketings of words with no repeated letters such as
$123 - 132 - 231 + 321 = [1,[2,3]]$.
We will also define  $\cL(A) \subset \Lie(A)$ to be the subspace of Lie
polynomials of {\it maximum length} ($\Lie(A)$  has Lie polynomials of different lengths). In particular, we write $\cL_{n-1}\equiv\cL(1,2,...,n-1) $  to be the linear span  of total bracketings of $1,...,n-1$. 

The left- and right-bracketings are surjective maps from $W$ onto $\Lie$ given by
\eqn\comb{
\ell[123\ldots n] := [[[1,2],3],\ldots,n]\,,\qquad
r[123\ldots n] := [1,[2,[3,\ldots,[n-1,n]\ldots]]].
}
Clearly
\eqn\ellPi{
\ell[Pi] = [\ell[P],i],\qquad r[iP] = [i,r[P]],
}
for a letter $i$ and word $P$. This inductively implies Baker's identity \Reutenauer
\eqn\ellPellQ{
\ell[P\ell[Q]] = [\ell[P],\ell[Q]].
}
Write $|P|$ for the length of a word $P$. Then $\ell[P]$ and $r[P]$ are related by
$\ell[P] = - (-1)^{|P|} r[\overline{P}]$,
where $\overline{P}$ is the reverse of $P$.

\newsubsec\dualsec The dual $\Lie^*$ of $\Lie$,   dual bases and their Kleiss-Kuijf relations

This section recalls the duality pairing between Lie polynomials and words defined up to shuffle products. This pairing is
central to the results of the paper.

Define first the Kronecker inner product, for words $P,Q \in W$, by
\eqn\pairing{
(P,Q) := \cases{1 & if $P=Q$;\cr
0 & otherwise.}
}
We will need the shuffle product on $W$, $\shuffle$, which is inductively defined by
\eqn\shuffleinduction{
(iP)\shuffle(jQ) := i (P\shuffle(jQ)) + j ((iP)\shuffle Q),
}
for letters $i,j$, and words $P,Q$. The base case is $i\shuffle j = ij + ji$. The expression $P\shuffle Q$ is sometimes referred
to as the sum over {\it ordered permutations} of $P$ and $Q$, preserving the ordering of the letters of $P$ and of $Q$. Ree's
theorem characterizes $\Lie$ in terms of the shuffle product. More precisely,
$\Gamma\in W$ is a Lie polynomial iff \Ree
\eqn\reestheorem{
(P\shuffle Q,\Gamma)=0
}
for all nonempty $P,Q\in W$.


If $Sh \subset W$ is the subspace spanned by all {\it proper shuffles}, $P\shuffle Q$ with $P,Q$ nonempty,
then Ree's theorem says that $\Lie = Sh^\perp$.  Thus the dual vector space to $\Lie$ is 
\eqn\whatliestar{
\Lie^* = W / Sh.
}
We will write elements of $\Lie^*$ as equivalence classes of the form $P+Sh$, for some $P\in W$. If two expressions, $P,Q\in W$ belong
to the same equivalence class, write $P\sim  Q$, i.e.
\eqn\Liestarequiv{
P\sim  Q\quad\Leftrightarrow\quad P=Q+Sh \quad \Leftrightarrow \quad P=Q + \sum_{A,B\neq \emptyset} A\shuffle B. \; 
}

\medskip

Dual bases of $\Lie$ and $\Lie^*$ can be obtained as follows. An element $P\in Lie^*$ can be represented by a particular
expression $P\in W$ in many ways. Likewise, a Lie polynomial $\Gamma\in\Lie$ is not uniquely written as a sum of complete
bracketings of words, because of the antisymmetry and Jacobi relations. It is therefore useful to find bases for $\Lie^*$ and
$\Lie$. 

A word $P$ is {\it Lyndon} if its first letter is also its smallest with respect to some fixed ordering of $\Bbb N$.\foot{$P$ is
assumed to have no repeated letters in our context. There is a more standard notion of Lyndon words for words that have repeated
letters \lothaire.} The set of dual Lie elements, $P+Sh\in\Lie^*$, where $P$ runs over all Lyndon words in $W$ is a basis of
$\Lie^*$ \Reutenauer.  Dually, the set of Lie monomials, $\ell[P]$, for all Lyndon words $P$ is a basis of $\Lie$. These two
bases are dual because, for two Lyndon words $P$ and $Q$, the smallest letter must come first in both words. But, for any letter
$i$,
\eqn\twolyndon{
(iP,\ell[iQ]) = (iP,iQ) = (P,Q),
}
because the only term in the word expansion of $\ell[iQ]$ that has $i$ at the beginning is $iQ$.

Restricting to $\cL_{n-1}^*$ this gives the basis of dual Lie elements $1Q+Sh$, for all the $(n-2)!$ permutations $Q$ of
$23...n-1$. Their linear independence follows from the fact that they are dual to the set of Lie monomials $\ell[1P]$ in $\cL$,
for $P$ a permutation of $23...n-1$ by \twolyndon.  Since $1Q$ spans $\Lie^*$, any $P+Sh\in\cL^*$ must have the following basis
expansion:
\eqn\Pbasisexpand{
P + Sh = \sum_Q (P, \ell[1Q]) 1Q + Sh,
}
where the sum is over all permutations, $Q$, of $23...n-1$.
The word expansions of $\ell[iP]$ and $r[Pi]$ have the following explicit formulas
\eqn\ellip{
\ell[iP] = \sum_{P\in X\shuffle Y} (-1)^\len{X} \bar X i Y\,,\qquad
r[Pi] = \sum_{P\in X\shuffle Y} (-1)^{|Y|} X i\bar Y\,.
}
These can be proved by verifying that they satisfy \ellPi.
Substituting $P=XiY$ into the explicit formula, \ellip, for the expansion of $\ell[iP]$, \Pbasisexpand\ implies that,
\eqn\schockerzero{
XiY \sim  (-1)^{|X|}i(\bar X \shuffle Y)\, ,
}
where $XiY$ is a permutation of $12...n-1$, and $i$ is some letter in $1,2,...,n-1$.\foot{Another way to
find \schockerzero\ is to prove the following identity at the level of words:
\eqn\ReeSchocker{
BiA - (-1)^{|B|}i(A\shuffle \bar B) =
- \sum_{XY=B\atop X\neq\emptyset}(-1)^{|X|} \bar X \shuffle (Y i A)\,.
}
This is stated in \Gauge\ and proven in \bandieraPC\ (see also equation (41) in \NovelliThibonToumazet).}
Setting $Y$ to be empty in \schockerzero\ gives for any word $P$ the special case
$P\sim -(-1)^{|P|}\bar P$.

 As explained in \BGBCJ, \schockerzero\ corresponds to the {\it Kleiss-Kuijf (KK) relations}
among color-ordered amplitudes \refs{\KKref,\KKproof}
\eqn\KKid{
\cA(X1Y,n) = (-1)^\len{X}\cA(1(\bar X\shuffle Y),n)
}
expressed now as a general
statement about $\Lie^*$ under the numerator homomorphism from the Lie polynomial amplitude \intrompn. Indeed
we will see that the KK relations follow  for any theory whose connected Feynman diagrams give rise to colour factors of the form \colourfactor.

We will also use basis expansions in $\cL_{n-1}$. If $\Gamma\in\cL_{n-1}$, then it has a basis expansion:
\eqn\Lyndonbasis{
\Gamma = \sum_Q ( 1Q,\Gamma)\,\ell[1Q]\,,
}
where the sum is over all permutations, $Q$, of $23...n-1$. For example, this gives
\eqn\onetwothreefour{
[[1,2],[3,4]] = \ell[1234] - \ell[1243],
}
which can also be verified directly using the Jacobi identity.

Finally, we will need to use the {\it adjoints} of $\ell$ and $r$, which we write as $\ell^*$ and $r^*$ :
\eqn\elladjoint{
(\ell^*(P),Q) = (P,\ell(Q)),\qquad (r^*(P),Q) = (P,r(Q)).
}
It follows from \ellPi\ that the adjoints can be computed recursively as
\eqnn\rhomaps
$$\eqalignno{
\ell^*(123 \ldots n) &= \ell^*(123 \ldots n-1)n - \ell^*(23 \ldots n)1,\qquad\ell^*(i):= i\,,&\rhomaps\cr
r^*(123 \ldots n) &= 1r^*(23 \ldots n) - n r^*(123 \ldots n{-}1),\qquad r^*(i):= i\,,\cr
}$$
Likewise, the explicit formulas \ellip\ are equivalent to the following formulas \michos
\eqn\elliP{
\ell^*(A) = \sum_{A=XiY} (-1)^\len{X}\,i (\bar X \shuffle Y)\,,\qquad
r^*(A) = \sum_{A=XiY}(-1)^{|Y|} (X \shuffle \bar Y)i\,.
}
Note that $\ell^*$ and $r^*$ are well-defined on $\Lie^*$ as $\ell^*(P\shuffle Q)=0=r^*(P\shuffle Q)$ for
nonempty $P,Q$.  This follows from $(\ell^*(P\shuffle Q), R)=(P\shuffle Q, \ell[R])$ which vanishes due to \reestheorem.

\ifig\figmonomial{The product of propagators ${1\over s_\Gamma}$ and the planar binary tree associated to the Lie
monomial $[[1,2],[3,4]]$ according to the definition \mangamma.}
{\epsfxsize=0.18\hsize\epsfbox{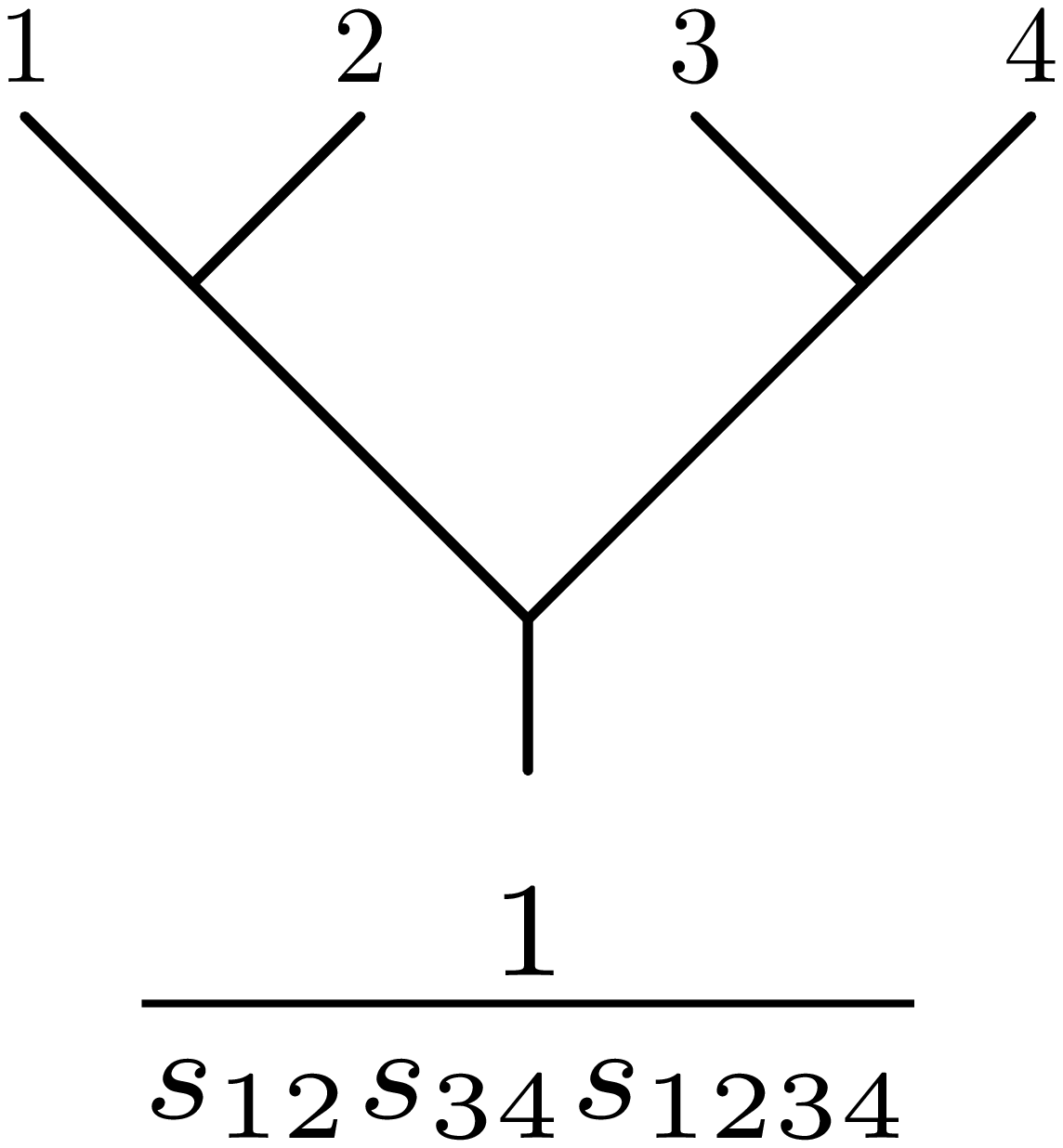}}

\newsubsec\mandbinsec Binary trees and colour structures

For any Lie monomial $\Gamma \in \cL_{n-1}$, there is a rooted binary tree $T_\Gamma$ \garsia. The inductive definition is as follows.
$T_{[1,2]}$ is the tree with two leaves, $1$ and $2$, connected by one vertex to the root. If $\Gamma = [\Gamma',\Gamma'']$,
then $T_\Gamma$ is the tree formed by connecting (or `grafting') the roots of $T_{\Gamma'}$ and $T_{\Gamma''}$ to make a new
vertex. Every pair of brackets in $\Gamma$ corresponds to a vertex in $T_\Gamma$. An example is shown in \figmonomial\ for
$\Gamma = [[1,2],[3,4]]$.

Fix $n$ elements of a Lie algebra: $t^a_i \in \cg$, for $i=1,...,n-1$. For any choice of the $t_i$, there is a map ${\bf t}: \cL_{n-1} \rightarrow \cg$. For a Lie monomial $\Gamma \in \cL_{n-1}$, ${\bf t}(\Gamma)$ is obtained by writing $\Gamma$ as a nested bracketing of $1,...,n-1$, and replacing every commutator bracket by the Lie bracket of $\cg$. If $\tr$ is the invariant inner product on $\cg$, then for every Lie monomial $\Gamma \in \cL_{n-1}$, the associated {\it colour factor} is
\eqn\colourfactor{
c_\Gamma = \tr ({\bf t}(\Gamma)t_n).
}
The replacement $\Gamma\mapsto c_\Gamma$ defines a homomorphism out of $\cL_{n-1}$; in particular $c_{-\Gamma} = - c_\Gamma$. If
$T_\Gamma$ is treated as a cubic Feynman diagram in a gauge theory with Lie group $\cg$, then (up to a sign) $c_\Gamma$ is the
colour factor of that diagram.\foot{The caveat about signs is because a cubic tree Feynman diagram in gauge theory is not just a
tree: the orientations at each vertex are important, and so the Feynman diagrams are sometimes drawn as quark graphs.}

\newsubsec\newman Mandelstam variables.

Massless scattering amplitudes are functions of external momenta, $k_i^\mu$, $i=1,...,n$, with $k_i\cdot k_i = 0$.The Mandelstam variable $s_{ij}$ is
\eqn\mandelstamvariable{
s_{ij} := k_i\cdot k_j.
}
For every subset $I\subset \Bbb N$ with at least two elements, write
\eqn\mansum{
s_I := \sum_{\{i,j\}\subset I} s_{ij}\,,\qquad
k_I^\mu := \sum_{i\in I} k_i^\mu,
}
so that $k_I\cdot k_J = \sum_{i\in I, j\in J} s_{ij}$.

Consider the Laurent ring with the variables $s_I$, for all subsets $I$ (with at least two elements).
Then let $\cS$ be the subring of this Laurent ring defined by the relation, \mansum. For example,
$1/(s_{12}+s_{23})$ is not a function in $\cS$. Whereas $1/(s_{12}+s_{23}+s_{13})$ is a function in $\cS$, because
\eqn\obvious{
\frac{1}{s_{12}+s_{23}+s_{13}} = \frac{1}{s_{123}}.
}
The functions in $\cS$ are the kind that arise in scattering amplitudes.

Fix a Lie monomial $\Gamma \in \cL_{n-1}$. When written as a nested bracket expression, each pair of brackets in $\Gamma$
defines a subset of $\{1,...,n-1\}$. If $I$ is a subset that appears like this, write $I\in \Gamma$, and define
\eqn\mangamma{
s_\Gamma := \prod_{I\in\Gamma} s_I.
}
For example, $s_{[[1,2],[3,4]]} = s_{12}s_{34}s_{1234}$, as in \figmonomial.

The inverse, $\frac{1}{s_\Gamma}$, is the {\it product of propagators} of the tree graph, $T_\Gamma$, associated to $\Gamma$,
including a propagator for the root of $T_\Gamma$. 

\newnewsec\BGsec Berends-Giele recursion and Lie polynomials

Berends-Giele (BG) recursion \BGpaper\ computes currents $J(P)^\mu$, labelled by permutations $P$, that can be used to find
the on-shell gauge theory tree amplitudes. In \FTlimit\ Berends-Giele recursion is applied to biadjoint scalar theory.  This led
to double colour-ordered currents $b(P|Q)$ labelled by two permutations $P$ and $Q$.\foot{In \FTlimit\ these were denoted
$\phi(P|Q)$.} The BG recursion from \FTlimit\ satisfied by $b(P|Q)$ induces a BG-like recursion for a parent object that was
denoted $b(P)$ in \PScomb. In this approach to Berends-Giele, $b(P)$ is a Lie polynomial with coefficients in the Mandelstam
variables. These $b(P)$, and the associated partial {\it Lie polynomial amplitudes}, are the basic subject of the subsequent
sections.

In the following we re-derive the results of \FTlimit\ and the appendix of \Gauge\ using fields with values in Lie polynomials.
Yang-Mills fields are similarly treated in appendix A.

\newsubsec\BGBS Berends-Giele recursion for  biadjoint scalar theories

Biadjoint scalars are scalar fields that take values in the tensor product of two Lie algebras $\Phi\in \cg\otimes \tilde {\cg} $.
Let these have structure constants $f^{abc}$ and $\tilde f ^{\tilde a \tilde b \tilde c}$ and invariant inner products for which we
take an orthonormal basis.  Then the Lagrangian is
\eqn\BS{
{\cal L}_{BS} = \frac12 \nabla_\mu \Phi_{a\tilde a} \nabla^\mu\Phi^{a\tilde a} + \frac 1{3!}  f_{abc}\tilde f_{\tilde a \tilde b \tilde c} \Phi^{a\tilde a}\Phi^{b\tilde b}\Phi^{c\tilde c}\, ,
}
where $\mu=1,\ldots,d$ is a space-time index and $a,b$ $\tilde a, \tilde b$ are Lie algebra indices for $\cg$,$\tilde {\cg}$.
The field equations are
\eqn\EOMBS{
\Box \Phi_{a\tilde a}= \frac12 f_{abc}\tilde f_{\tilde a \tilde b \tilde c} \Phi^{b\tilde b}\Phi^{c\tilde c}\, .
}

Consider the field $\Phi(x) \in \Lie\otimes \Lie$ rather than $\cg\otimes \tilde {\cg}$ now satisfying
\eqn\EOMBSLP{
\Box \Phi= \frac12  [\![ 
\Phi,\Phi]\!] \, ,
}
where $[\![~,~]\!]$ is the symmetric operation:
\eqn\bbracket{
[\![\Gamma_1\otimes \tilde\Gamma_1,\Gamma_2\otimes \tilde\Gamma_2]\!]:=[\Gamma_1,\Gamma_2]\otimes [\tilde\Gamma_1,\tilde\Gamma_2]\, .
}

Inverting the wave operator in \EOMBS\ gives the recursion
\eqn\EOMBSit{
\Phi_{n}= \Phi_1+ {\rm proj}_{\Lie_{\leq n}\otimes \Lie_{\leq n}}\Box ^{-1} \frac12 [\![ 
\Phi_{n-1},\Phi_{n-1}]\!] 
\, ,
}
for a perturbative solution in the form of fields $\Phi_{n}\in \Lie_{\leq n}\otimes \Lie_{\leq n}$, where $\Lie_{\leq n}$ is the subspace spanned by Lie monomials $\Gamma$ with length $|\Gamma|\leq n$.
The iteration is seeded by the following homogeneous solution to $\Box \Phi_1=0$:
\eqn\seededby{
 \Phi_1=\sum_{j\in{\Bbb N}} e^{ik_j\cdot x}\, j\otimes j\, \in \Lie\otimes \Lie\, .
 }
By our definition of $\Lie$, the letters appearing in monomials in $\Phi_n$ are distinct.\foot{In the perturbiner approach
this is enforced by introducing nilpotents into each seed solution.}  The coefficient of a word $P$ in $\Phi_n$ has $x$
dependence $e^{ik_P\cdot x}$. The inverse wave operator, $\Box^{-1}$, acts on such a term to give $1/k_P^2=1/s_P$.
By construction, $\Phi_n$ is symmetric in its two factors:
\eqn\Phisym{
\Phi_n=\sum_j \phi_j \Gamma_j\otimes \Gamma_j\,,
}
for some coefficients $\phi_j$,  and Lie monomials $\Gamma_j$. It follows that, for a word $P$, pairing with the left
or right factor of $\Phi_n$ gives $(P,\Phi_n)_L = (P,\Phi_n)_R$. So write
\eqn\PhiP{
\Phi(P) := (P,\Phi)_L = (P,\Phi)_R.
}
Define now the Berends-Giele currents to be
\eqn\bdef{
b(P)= e^{-i k_P\cdot x} \Phi(P) \quad \in \quad \Lie_s :=\Lie \otimes {\cal S}\, ,
}
i.e., Lie polynomials multiplied by rational functions of the Mandelstams, where the factor of $e^{-ik_P\cdot x}$ removes the $x$ dependence.  

\proclaim Proposition \PScomb. The Berends-Giele currents satisfy
\eqn\BMap{
b(P) = {1\over s_P}\sum_{XY=P}[b(X),b(Y)]\,, \qquad b(i) = i\, ,
}
where the sum is over all deconcatenations, $P=XY$, of $P$. 

\proof This follows by pairing \EOMBSit\ with $P$ and using the identity 
\eqn\identitycheck{
(P, [\Gamma_{j_1},\Gamma_{j_2}])\otimes [\Gamma_{j_1,}\Gamma_{j_2}]
=\sum_{XY=P} [(X,\Gamma_{j_1})\Gamma_{j_1}, (Y,\Gamma_{j_2})\Gamma_{j_2}] + [(X,\Gamma_{j_2})\Gamma_{j_2}, (Y,\Gamma_{j_1})\Gamma_{j_1}].
}
which together with the symmetry of $\Phi$ gives
\eqn\BCPhifin{
\frac{1}{2} [\![ \Phi,\Phi]\!](P) = \sum_{XY=P}[ \Phi(X) ,\Phi(Y)].
}
which gives   \BMap. \qed

For example, the recursion \BMap\ gives
\eqnn\bexampone
$$\eqalignno{
b(12) &= {[1,2]\over s_{12}},&\bexampone\cr
b(123) &= {[[1,2],3]\over s_{12}s_{123}}  + {[1,[2,3]]\over
s_{23}s_{123}}\,,\cr
b(1234) &= {[ [ [ 1 , 2 ] , 3 ] , 4 ] \over s_{12} s_{123} s_{1234}}
+  {[ [ 1 , [ 2 , 3 ] ] , 4 ] \over s_{123} s_{1234} s_{23}}
+  {[ [ 1 , 2 ] , [ 3 , 4 ] ] \over s_{12} s_{1234} s_{34}}
+  {[ 1 , [ [ 2 , 3 ] , 4 ] ] \over s_{1234} s_{23} s_{234}}
+  {[ 1 , [ 2 , [ 3 , 4 ] ] ] \over s_{1234} s_{234} s_{34}}\,.
}$$
Since $\Phi \in \Lie\otimes \Lie$, Ree's theorem implies that 
\eqn\bMapshuffle{
b(R\shuffle S) = 0\,,
}
for nonempty $R,S$.  Thus  $b: P \mapsto b(P)$ defines a map
\eqn\bMaphere{
b: \Lie_s^* \rightarrow \Lie_s\,.
}

To obtain  an $n$-particle  amplitude from a Berends-Giele current,  remove the last propagator
by multiplying by $s_{12...n-1}$, and imposing momentum conservation this becomes $k_n^2$ which we
then set to zero. Thus we  define the $\Lie$-valued ``amplitude'' by
\eqn\Lieamp{
m(P,n)=\lim_{s_{P}\rightarrow 0}  s_{P}\,  b(P)  \; \in \Lie_\cS,
}
for permutations $P$ of $12...n-1$ (although \bMapshuffle\ in fact gives $P\in \cL_{n-1}^*$.)

\newsubsec\pbtsec The tree diagram expansion of $b(P)$

The Feynman diagrams of biadjoint scalar theory are trivalent trees. Berends-Giele recursion
generates the Feynman graphs with  one off-shell  leg decorated with  an off-shell propagator,
with the remaining legs on-shell. So $b(P)$ can be written as a sum over binary trees as follows. Write
\eqn\btreeformula{
b := \sum_{\Gamma} \frac{\Gamma\otimes\Gamma}{ s_\Gamma} \quad \in \quad \Lie\otimes \Lie,
}
where the sum is over all Lie monomials (up to sign), and where $ s_\Gamma$ is the product of
variables $s_I$ defined in \mangamma.  Then pairing $b$ with a word $P$ gives
\eqn\bmaptwo{
b(P) = \sum_\Gamma \frac{(P,\Gamma)\Gamma}{s_\Gamma},
}
which is a sum over all planar rooted trees with respect to the ordering $P$.

\ifig\figBGfour{The Catalan expansion $b(1234)$ from \BMap. Viewed as cubic graphs and removing the overall propagator $1/s_{1234}$,
they correspond to the expansion of a color-ordered five-point tree amplitude $A(12345)$ \BCJ. Note the leg $5$ does not enter
in the Lie elements in the numerators and that the root is unlabeled. By labelling the root and assigning leg $5$ to the
Catalan expansion of $b(5)$ one recovers the free Lie algebra correspondence \BGpbt\ for the case $n=5$.}
{\epsfxsize=0.88\hsize\epsfbox{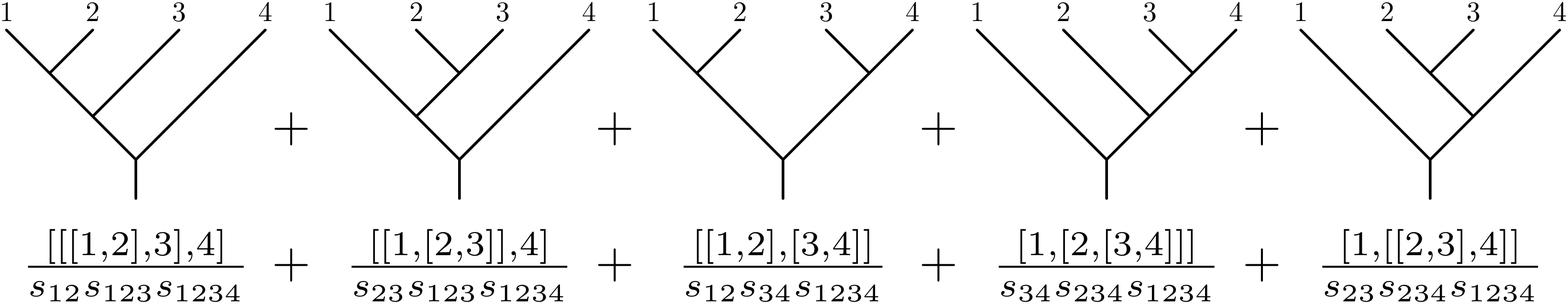}}

The following Lemma shows that \bmaptwo\ is indeed the solution to the Berends-Giele recursion discussed in \S\BGBS.

\proclaim Lemma. The formula in \bmaptwo\ satisfies \BMap.

\proof For any Lie monomial $\Gamma$ there are $\Gamma_1$ and $\Gamma_2$  so that $\Gamma=[\Gamma_1,\Gamma_2]$, and these monomials are unique up to sign. So, for a fixed Lie monomial $\Gamma$,
\eqn\provingthesame{
\frac{(P,\Gamma)\Gamma}{s_\Gamma}= \frac{(X,\Gamma_1)(Y,\Gamma_2) [\Gamma_1,\Gamma_2]}{s_P s_{\Gamma_1}s_{\Gamma_2}} - \frac{(Y,\Gamma_1)(X,\Gamma_2) [\Gamma_1,\Gamma_2]}{s_P s_{\Gamma_1}s_{\Gamma_2}},
}
where $P=XY$ and $|X|=|\Gamma_1|$, $|Y|=|\Gamma_2|$. Summing over all Lie monomials (up to sign), $\Gamma_1$ and $\Gamma_2$, that have length strictly smaller than $|P|$ gives \BMap. \qed

Regarding $b$ as a map from $\Lie_\cS^*$ to $\Lie_\cS$ it follows from \btreeformula\ that $b$ is self-adjoint:
\eqn\biadjointbg{
b(P|Q):= (P,b(Q)) = (b(P),Q)=b(Q|P)\, ,
}
for any $P,Q\in \Lie_\cS^*$.
Since $b(P)$ is a Lie polynomial it can be expanded in a basis as
\eqn\bphiell{
b(P) = \sum_R b(P|iR)\ell[iR],
}
where $i$ is some letter in $P$ and we used \Lyndonbasis.

The usual biadjoint scalar partial tree amplitude is \FTlimit
\eqn\BGpbt{
m(Pn,Qn))=:  \lim_{s_{P}\rightarrow 0}s_{P} b(P|Q) = (Q,m(Pn))\,.
}
We will see in \S\Atheorysec\ how $b(P)$ can be dressed with BCJ numerators to give BG currents and hence amplitudes
for certain coloured theories.

\newnewsec\secbcj A Lie bracket for tree-level scattering amplitude relations

This section introduces a Lie bracket on $\Lie^*_\cS$, and uses it to prove the fundamental BCJ relation. Following
\PScomb, this Lie bracket is then used to define a generalized KLT matrix in \S\secklt\ and to prove the identities conjectured
in \PScomb.

\newsubsec\newLie The $S$-bracket

The {\it S-map} was introduced in \refs{\EOMbbs,\BGBCJ} to express the BCJ relations for super-Yang--Mills amplitudes from its
action on Berends-Giele currents $M_P$ from \nptMethod. It was abstracted to a map
acting on words in \PScomb\ and the off-shell BCJ relations for $b(P)$ was conjectured, but no general proof was given. Here we
will see that the S-map defines a Lie bracket on $\Lie^*$ and this will lead to a proof of the fundamental BCJ
relations.

\proclaim Definition ($S$ bracket). Define a multilinear pairing $\{~,~\}: \cL^*\otimes\cL^* \rightarrow \cL^*$ by \PScomb
\eqn\curlydef{
\{P,Q\} := r^*(P)\star \ell^*(Q),
}
where $r^*$ and $\ell^*$ are defined in \rhomaps\ and
\eqn\hash{
Ai\star jB:= s_{ij}\, AijB
}
for words $A, B$ and letters $i, j$.
Equivalently, for $P,Q\in \Lie^*$, equation \elliP\ gives \EOMbbs
\eqn\curlydeftwo{
\{P,Q\} = \sum_{XiY=P\atop RjS=Q} s_{ij}\,(X\shuffle \a(Y))ij(\a(R)\shuffle S).
}\par
\noindent Alternatively, a recursive definition is given by
\eqnn\smapdef
$$\eqalignno{
\{iAj,B\}&=i\{Aj,B\}-j\{iA,B\},&\smapdef\cr
\{B,iAj\}&=\{B,iA\}j-\{B,Aj\}i,\cr
\{i,j\} &= s_{ij}\,ij
}$$
as can be verified from the explicit form of the adjoint maps $\ell^*$ and $r^*$ in \elliP.
For example,
\eqnn\curlydefex
$$\eqalignno{
\{1,2\} &=s_{12}12,  &\curlydefex\cr
\{1,23\} &= s_{12} 123 - s_{13} 132,\cr
\{12,3\} &= s_{23} 123 - s_{13} 213,\cr
\{1,234\}&= s_{12} 1234 - s_{13} 1324 - s_{13} 1342 + s_{14} 1432,\cr
\{123,4\}&= s_{34} 1234 - s_{24} 1324 - s_{24} 3124 + s_{14} 3214,\cr
\{12,34\} &= s_{23}1234 - s_{24}1243 - s_{13} 2134 + s_{14}2143.
}$$
Given that the adjoints $r^*$ and $\ell^*$ annihilate proper shuffles, the
definition \curlydef\ manifestly satisfies $\{A\shuffle B,C\}=0$.

The $S$ bracket is antisymmetric in $\Lie^*$. Indeed, by \curlydeftwo,
\eqn\curlydefasym{
\{Q,P\} = \sum_{XiY=P\atop RjS=Q} s_{ij}\,(R\shuffle\a(S))ji(\a(X)\shuffle Y) \sim  - (-1)^{Q+P}\overline {\{P,Q\}},
}
because, in $\Lie^*$, $X$ is shuffle equivalent to $-(-1)^X\overline X$. Using \Liestarequiv\ this means that,
\eqn\curlyasym{
\{P,Q\}\sim  -\{Q,P\}.
}
In fact, we show in \S\absKLTsec\ that the S bracket is a Lie bracket on $\Lie_\cS^*$, and so it also satisfies the Jacobi identity.

\newsubsec\BCJsec The BCJ amplitude relations

This section proves the main property of the $S$ bracket which amounts to a generalization of the well-known {\it
off-shell fundamental BCJ relation} \Du. Following \EOMbbs, we call the relations implied by the $S$ bracket {\it BCJ amplitude relations}. The on-shell
identities follow directly from the off-shell relations, as is explained at the end of this section.

\proclaim Proposition. For $P,Q\in \Lie^*$, the $S$-bracket satisfies
\eqn\bigclaim{
b(\{P,Q\}) = [b(P),b(Q)]\,,
}
i.e., $b$ maps the $S$ bracket to the Lie bracket.

The proposition is proved in appendix \mainapp. It is interesting to observe that the property \bigclaim\ mimics the identity
obeyed by the Poisson bracket $\{,\}$ of Hamiltonian vector fields $X_f$: $X_{\{f,g\}} = [X_f,X_g]$ for functions $f$ and $g$.

\proclaim Corollary (Fundamental off-shell BCJ  relations \Du). Taking $P=i$, a single letter, we obtain
\eqn\bcjrelOS{
b(\{i,Q\}) :=\sum_{Q=XjY} s_{ij} b(ij\alpha (X)\shuffle Y )= [i,b(Q)],
}
where $\alpha(X)=(-1)^{|X|}\bar X$.

The expression in \bcjrelOS\ can also be written as
\eqn\ordinaryfundbcj{
\{i,Q\} \sim \sum_{Q=XY} k_i\cdot k_Y XiY 
}
To see this, use \Pbasisexpand\ to write the RHS \ordinaryfundbcj\ in the basis of words beginning with the letter $i$,
\eqn\ordinarytocurly{
\sum_{Q=XY} (k_{i}\cdot k_{Y}) \, XiY \sim \sum_{Q=XY} (k_{i}\cdot k_{Y}) \, i (\alpha(X)\shuffle Y)
}
Further manipulations give
\eqnn\moreordinary
$$\eqalignno{
RHS & = \sum_{Q=XjY} (k_{i}\cdot k_{jY}) \, ij (\alpha(X)\shuffle Y) - (k_{i}\cdot k_{Y})\,ij (\alpha(X)\shuffle Y),&\moreordinary\cr
& = \sum_{Q=XjY} s_{ij} \,  ij (\alpha(X)\shuffle Y),
}$$
where we use the property \shuffleinduction\ of the shuffle product.

\proclaim Corollary (BCJ relations). The tree-level partial amplitudes \Lieamp\ satisfy
\eqn\bcjrel{
m(\{P,Q\},n) = 0\,,
}
where $P$ and $Q$ are words that partition $1,2,...,n-1$ into two parts.

\proof By the definition \Lieamp,
\eqn\vanishing{
m(\{P,Q\},n) =\lim_{s_{PQ}\rightarrow 0} s_{PQ} b(\{P,Q\})= \lim_{s_{PQ}\rightarrow 0}s_{PQ}[b(P),b(Q)] = 0\,.
}
The last term vanishes because neither $b(P)$ nor $b(Q)$ has a $1/s_{PQ}$ pole.\qed

The original fundamental BCJ relations are 
\refs{\fundamentalBCJ,  \KLTmatrixrefII}
\eqn\fund{
m(\{i,Q\}n) = \sum_{Q=RS} (k_i\cdot k_S)m(RiSn) = 0,
}
which follows from \ordinaryfundbcj. For example, by \curlydefex, \BCJ
\eqnn\bcjfour
$$\eqalignno{
0&= m(\{1,23\}4) = s_{12} m(1234) - s_{13} m(1324),&\bcjfour\cr
0&=m(\{1,234\},5) = s_{12} m(12345) - s_{13} m(13245) - s_{13} m(13425) + s_{14} m(14325).
}$$
But, as observed in \refs{\EOMbbs,\BGBCJ}, \bcjrel\ also implies other BCJ relations, such as
\eqn\bcjfive{
0=m(\{12,34\},5) = s_{23} m(1234,5) - s_{13} m(2134,5) - s_{24} m(1243,5) + s_{14} m(2143,5),
}
while similar formulas using the shuffle product appear in \refs{\monodVanhove,\monodStie,\dushuffle}.
The BCJ relations for Yang--Mills theory were first proven from the field-theory limit
of string theory in \monodVanhove\ and \monodStie. By now these relations have been proven for
a variety of theories at tree-level. See the recent review \JJreview\ and references therein.

\newnewsec\secklt The KLT map

This section introduces a canonical KLT map $S: \Lie_\cS \rightarrow \Lie^*_\cS$. $S$ is the inverse of $b$, and this implies
that the $S$-bracket is a Lie bracket. When $S$ is written out in terms of a pair of bases for $\Lie$ and $\Lie^*$, the matrix
elements of $S$ are the generalized KLT matrix $S(P|Q)$ proposed in \PScomb. We prove the properties conjectured in \PScomb\ as
well as additional ones.

\newsubsec\absKLTsec The KLT map

Let $\Gamma$ be a Lie monomial, and write it as a nested bracketing. Let $\{\Gamma\} \in \Lie^*$ be obtained by replacing every
commutator $[~,~]$ in the bracketed expression of $\Gamma$ with a $\{~,~\}$. This is well defined because the $S$-bracket is antisymmetric. By nested applications of \bigclaim, it follows from the proposition in the previous section that
for any Lie monomial $\Gamma$,
\eqn\curlylabla{
\Gamma = b(\{\Gamma\}).
}
\noindent For example,
\eqnn\bexam
$$\eqalignno{
[[1,2],[3,4]]&= b(\{\{1,2\},\{3,4\}\})&\bexam\cr
&= s_{12}s_{34}\bigl(s_{23}b(1234) - s_{24}b(1243) - s_{13} b(2134) + s_{14}b(2143)\bigr).
}$$
where we used $\{\{1,2\},\{3,4\}\}=s_{12}s_{34}\{12,34\}$, the example \curlydefex\ and the relations
among Mandelstam invariants.

Define the KLT map $S:\Lie_\cS\rightarrow \Lie^*_\cS$ by  
\eqn\SGamma{
 S:\Gamma \mapsto \{\Gamma\},
 }
for Lie monomials $\Gamma$. As written, it is not obvious that $S$ is well defined. The reason \SGamma\ is well defined is that it turns out that $S$-bracket is Lie. This is shown in the proof of the following main proposition:

\proclaim Proposition. The maps $b:\Lie^*\rightarrow\Lie$ and $S:\Lie\rightarrow \Lie^*$ are inverses. In particular, $b$ is invertible.

\proof 
Choose an ordering of $A$, which defines dual Lyndon bases of $\Lie^*(A)$ and $\Lie(A)$, as in \S\dualsec. Then define a map, $S'$:
\eqn\Spropergooddef{
S' : \ell(a) \mapsto \{\ell(a)\},
}
for monomials $\ell(a)$ in the given basis of $\Lie(A)$. We show that (i) $S'$ and $b$ are inverse, and (ii) that the $S'$ in \Spropergooddef\ is the map, $S$, in \SGamma. This proves that \SGamma\ is well-defined. 

By \curlylabla,
\eqn\bInvs{
b(S'(\Gamma)) = \sum_{P\in Basis} (\Gamma,P) b(\{\ell(P)\}) = \sum_{P\in Basis} (\Gamma,P) \ell(P) = \Gamma.
}
Conversely,
\eqn\Sbinvs{
S'(b(P)) = \sum_{Q\in Basis} (b(P),Q) \{\ell(Q)\}.
}
Expanding $\{\ell(Q)\} \in \Lie^*$ in the given basis gives (by \Pbasisexpand),
\eqn\Sbinvsstep{
\{\ell(Q)\} = \sum_{R\in Basis} (\{\ell(Q)\},\ell(R)) R.
}
But notice that
\eqn\bInvsSw{
(\{\Gamma\},\Gamma') = (\{\Gamma\},b(\{\Gamma'\})) =(b(\{\Gamma\}),\{\Gamma'\}) = (\Gamma,\{\Gamma'\}),
}
by the self-adjointness of $b$. Combining \Sbinvsstep\ and \bInvsSw\ gives that
\eqn\SInvs{
S'(b(P)) = \sum_{Q,R \in Basis} b(P,Q) (\ell\{Q\},\ell(R)) R = \sum_{R\in Basis} (P,\ell(R)) R = P.
}
So $b$ is invertible, with inverse $S'$. But $b(\{\Gamma\}) = \Gamma$, by \curlylabla. So $S'(\Gamma) = \{\Gamma\}$.
\qed

Notice that the self-adjointness of $b$ implies (as in \bInvsSw) that the KLT map, $S$, is self adjoint:
\eqn\Sself{
(S(\Gamma),\Gamma') = (\Gamma,S(\Gamma')).
}

\proclaim Corollary.  The $S$-bracket is Lie.

\proof
Given that $S$ and $b$ are inverse, \bigclaim\ shows that $\{~,~\}$ is the pull back
of $[~,~]$ from $\Lie$ to $\Lie^*$ by $b^{-1}$.  i.e.  $\{~,~\}$  is skew and satisfies the Jacobi relation.
\qed

\newsubsec\genKLTsec The generalized and standard KLT matrix

This section finds explicit formulas for the KLT map, $S$, in terms of the matrix elements of $S$. In the next section, it is
shown how these reduce to the standard formula for the KLT matrix elements.

If $P_i,\Gamma_i$ is a pair of dual bases for $\Lie^*$ and $\Lie$, then
\eqn\Sinbases{
S: \Gamma \mapsto \sum_{i,j} (P_i,\Gamma) (\{\Gamma_i\}, \Gamma'_j) P'_j.
}
In other words, the matrix elements of $S$ are
\eqn\Sonetwo{
S(\Gamma_1,\Gamma_2) = (\{\Gamma_1\}, \Gamma_2),
}
for two Lie monomials $\Gamma_1, \Gamma_2$.

For the rest of this section, we again take pairs of dual Lyndon bases: $P\in \Lie^*$ for every Lyndon word $P$, and $\ell[P] \in
\Lie$ for every Lyndon word $P$. In these bases, the matrix elements, \Sonetwo, are labelled by Lyndon words. So define the {\it
generalized KLT matrix} (introduced in \PScomb)\foot{Alternative formulas for the generalized KLT matrix arise by choosing any
pair of dual bases.  E.g.\ one can define $S^{r} (P,Q) := (r\{P\},r(Q))$, but as $P, Q \in \Lie$, by Lemma 1.7 of \Reutenauer,
$r\{P\}=-\ell\{\a(P)\}$ and $r(P)=-\ell(\a(P))$ so $S^{r}(P|Q)=S^{\ell}(\bar P|\bar Q)$. These alternative choices explain
redefinitions w.r.t reversal of permutations $P\to\bar P$, see e.g.\ footnote 10 in \Polylogs.} to be these matrix
elements:
\eqn\defgenklt{
S^\ell (P|Q) := (\ell\{P\},\ell[Q]).
}
In this formula, we introduce the notation $\ell\{P\}$ for $\{\ell[P]\}$.\foot{In \PScomb\ the
map $\ell\{A\}$ was denoted $\sigma(A)$.} This is given explicitly by:
\eqn\curlyellp{
\ell\{12...n\} := \{...\{1,2\},...,n\}.
}
For example,
\eqnn\curlyellpexample
$$\eqalignno{
\ell\{12\} &= s_{12} 12,&\curlyellpexample\cr
\ell\{123\} &= s_{12}s_{23} 123 - s_{12} s_{13} 213,\cr
\ell\{1234\} &=
       + s_{12} s_{23} s_{34} 1234
       - s_{12} s_{23} s_{24} 1324
       - s_{12} s_{13} s_{34} 2134
       + s_{12} s_{13} s_{14} 2314\cr
&\quad{}       - s_{12} s_{13} s_{24} 3124
       - s_{12} s_{23} s_{24} 3124
       + s_{12} s_{13} s_{14} 3214
       + s_{12} s_{14} s_{23} 3214.
}$$
Some example entries of the generalized KLT matrix are
\eqn\genkltexamples{
\eqalign{
S^\ell(12|12) &= s_{12},\cr
S^\ell(12|21) &= - s_{12},\cr
}\qquad
\eqalign{
S^\ell(123|123) &= s_{12}(s_{13}+s_{23}),\cr
S^\ell(312|123) &= - s_{12}s_{13}.
}\qquad
\eqalign{
S^\ell(132|123) &= s_{12}s_{13},\cr
S^\ell(321|123) &= s_{12}s_{23}.
}}
The generalized KLT matrix is the matrix of coefficients that arises in the basis expansion of $\ell\{P\}$:
\eqn\curlybasisexpand{
\ell\{P\} \sim  \sum_{Q\in Basis} (\ell\{P\}, \ell[Q]) Q=\sum_{Q\in Basis} S^\ell(P|Q)Q ,
}
as in \Sbinvsstep.
As pointed out in \PScomb, the generalized KLT matrix $S^\ell(P|Q)$ can be defined for {\it any} two words $P,Q$, instead of
restricting $P,Q$ to a set of Lyndon words.

It follows from \bInvsSw\ that the generalized KLT matrix is symmetric:
\eqn\itis{
S^\ell(P|Q) = S^\ell(Q|P).
}
Moreover, \bInvs\ implies that the generalized KLT matrix satisfies
\eqn\genkltrelation{
\ell[P] = \sum_{Q\in Basis} S^\ell(P|Q)\, b(Q).
}
Or, equivalently,
\eqn\cortwo{
(R,\ell[P]) = \sum_{Q\in Basis} S^\ell(P|Q)\, b(Q|R),
}
and so, also:
\eqn\corone{
b(P|Q) = \sum_{X,Y \in Basis} \, b(P|X)\, S^\ell(X|Y) \, b(Y|Q).
}
We now show that the standard KLT matrix
arises from the restriction \PScomb
\eqn\firsti{
S(P|Q)_i := S^\ell(iP|iQ),
}
for some fixed letter $i$, called the `fixed leg'.
An explicit formula for $S(P|Q)_i$ was given in
\refs{\oldMomKer,\KLTmatrixrefI}. This explicit formula can be
recovered\foot{An alternative explicit derivation using \firsti\ is given in appendix C, where the steps are closer in spirit to the
argument in \Du.}
from \firsti\ using the following recursion, originally conjectured in \NLSM.

\proclaim Lemma. The standard KLT matrix \firsti\ can be recursively computed using
\eqn\carlosrec{
S(Aj|BjC)_i = k_j\cdot k_{iB}\, S(A|BC)_i,
}
where $i$ is the fixed leg. The recursion is seeded by $S(\emptyset|\emptyset)_i := 1$.

\proof
By definition, $\ell\{Aj\} = \{\ell\{A\},j\}$. Using the formula \ordinaryfundbcj\ for $\{P,j\}$,
\eqn\ellcAiexp{
\ell\{Aj\} = \sum_{X,Y} k_j\cdot k_X\,(XY,\ell\{A\}) XjY.
}
Since $S(Aj|BjC)_i = (\ell\{iAj\},\ell(iBjC))$, it follows from \ellcAiexp\ that
\eqn\carlosrecone{
S(Aj|BjC)_i = \sum_{X,Y} k_j\cdot k_X (XY,\ell\{iA\}) (XjY,\ell(iBjC)).
}
Expanding $\ell\{iA\}$ in a basis, as in \curlybasisexpand,
\eqn\ellmAex{
\ell\{iA\} = \sum_P S(A|P)_i\, iP.
}
So \carlosrecone\ becomes
\eqn\carlosrectwo{
S(Aj|BjC)_i = \sum_{X,Y,P} k_j\cdot k_X S(A|P)_i (XY,iP) (XjY,\ell(iBjC)).
}
The only contributions in the sum come from words, $X=iX'$, that begin with the letter $i$.
Expanding $\ell(iBjC)$, one sees that $(iX'jY,\ell(iBjC)) = \delta_{X',B}\delta_{Y,C}$. \qed

The identity, \cortwo, together with the definition \firsti\ is a purely algebraic proof that the standard KLT
matrix is the inverse to the biadjoint scalar BG double current:
\eqn\invKLT{
\sum_R S^\ell(P|R)_i b(iR|iQ) = \delta_{P,Q}\,.
}
This is in accord with the discussions of \DPellis\ and \FTlimit.

\newsubsec\KLTmomsec The momentum kernel  and the KLT gravity formula

The $(n-2)!$ version of the KLT relation, as given by \refs{\KLTmatrixrefI,\KLTmatrixrefII}, is
\eqn\ntwoKLT{
M_n = \lim_{s_{1P}\rightarrow 0} \sum_{P,Q\in S_{n-2}} {1\over s_{1P}}A(1Pn)S(P|Q)_1 {\tilde A}(1Qn),
}
where $\tilde A(1Qn)$ is obtained from its left-moving counterpart by replacing the polarization vectors, $e_i^\mu$,
by some other set of polarization vectors, $\tilde e_i^\mu$. It takes some effort to see directly that the
limit on the RHS of \ntwoKLT\ is well defined. The cancellation of the $1/s_{1P}$ pole on the RHS is easily
understood in terms of the above discussion of $S(P|Q)_1$ in \S\absKLTsec.

For example, starting from \invKLT, we recover the well known fact that the KLT matrix \defgenklt\ annihilates the on-shell biadjoint scalar amplitudes,
\eqn\momKern{
\sum_Q S^{\ell}(1P|1Q)m(1Q,n|R,n) = 0\,,
}
where $m(1Q,n|R,n) = \lim_{s_{iP}\rightarrow 0} s_{iP} b(1Q|R)$. The expression vanishes because there is no $1/s_{iP}$ pole in \invKLT.
This is a remarkable identity, because the components $S(P,Q)_i$ do not have  $s_{iP}$ as a factor. On the other hand,
$b(iP,iQ)$ {\it does} have a $1/s_{iP}$. This cancellation is the key reason why \ntwoKLT\ is well defined.

In the rest of this section, we explain how a Lie polynomial version of \ntwoKLT\ follows immediately
from the main result in \S\absKLTsec. We have already shown, in \corone, that the KLT matrix satisfies the following off-shell, $(n-2)!$ KLT relation
\eqn\coronetwo{
b(P|Q) = \sum_{R,U} b(P|iR)S(R|U)_i b(iU|Q).
}
To obtain a relation of the form \ntwoKLT, write 
\eqn\KLTfreeA{
A(1P,n) := s_{1P} b(1P) = s_{1P}\sum_{R\in S_{n-2}}
b(1P|1R)\ell[1R],
}
for the Lie polynomial `precursor' of a gauge theory amplitude. And write
\eqn\KLTfreeM{
M_n =  \sum_{P,Q\in S_{n-2}} s_{1P} \,\ell[1P]b(1P|1Q)\ell[1Q]
}
for the precursor of a gravity amplitude. Notice that, by \bphiell, \KLTfreeM\ is equivalently written as
\eqn\bn{
M_n = s_{12...n-1} \sum_{|\Gamma| = n-1} \frac{\Gamma\otimes\Gamma}{s_\Gamma}.
}
Then \coronetwo\ implies the following KLT relation
\eqn\KLTfree{
M_n = \lim_{s_{1P}=0}\sum_{P,Q\in S_{n-2}} {1\over s_{1P}}\, S(P|Q)_1\, m(1Pn)\, \otimes  \, m(1Qn).
}
All the double poles from the right-hand side of \ntwoKLT\ are manifestly absent in its free
Lie algebra version \KLTfree\ due to the generalized KLT matrix property \coronetwo. The only poles in \coronetwo\ are
those that appear in $b(1P|1Q)$.

Our results also make it clear why the existence of a `BCJ form' for gravity is equivalent to the KLT relation.
As an example, the four-point ``gravity'' amplitude has the following double copy expansion:
\eqnn\gravfour
$$\eqalignno{
M_4 &= s_{123}\Bigl(b(123|123)\tilde\ell[123]\ell[123] + b(123|132)\tilde\ell[123]\ell[132] &\gravfour\cr
&\qquad{}+b(132|123)\tilde\ell[132]\ell[123] + b(132|132)\tilde\ell[132]\ell[132]\Bigr)\cr
&={[[1,2],3]\otimes [[1,2],3] \over s_{12}} + {[[1,3],2]\otimes [[1,3],2]\over s_{13}}
+ {[1,[2,3]]\otimes [1,[2,3]]\over s_{23}},
}$$
where we used $\ell[123]-\ell[132] = [1,[2,3]]$.


\newsubsubsec\Liealgsec Generalized Jacobi identities

The main property of the bracket the $S$-bracket is that
\eqn\bracketagain{
b(\{P,Q\}) = [b(P),b(Q)].
}
Since $b$ and $S$ are inverse, this is equivalent to
\eqn\bracketagain{
\{P,Q\} = S([b(P),b(Q)]).
}
The bracket $\{P,Q\}$ is polynomial in the Mandelstam variables, whereas $S([b(P),b(Q)])$ is naively a rational function of the
Mandelstam variables. The cancellation of the poles in $[b(P),b(Q)]$ by the KLT matrix is not at all obvious from its concrete
formula.

As observed in \S\absKLTsec, \bracketagain\ implies that the $S$-bracket is Lie, and so satisfies
\eqn\twolie{
\{1,2\}\sim  - \{2,1\}\,\qquad \{\{1,2\},3\}+\{\{2,3\},1\}+\{\{3,1\},2\} \sim  0,
}
as identities in $\Lie^*_\cS$. It follows that the $S$-bracket satisfies generalized Jacobi identities \Reutenauer
\eqn\curlybracketshuffle{
\ell\{PiQ\}\sim  -\ell\{i\ell[P]Q\}
}
which can be deduced from two applications of \ellPellQ. This makes it clear that the definition
of $S^\ell(P,Q)$, given in \defgenklt\ also satisfies such generalized Jacobi identities in $P$ and $Q$ \PScomb,
\eqn\sPQJacobi{
S(XiY|Q) = - S(i\ell[X]Y|Q).
}
Notice that this property of the generalized KLT matrix has no analog for the standard KLT matrix,
$S(XiY|Q)_j$, because of the fixed leg $j$.\foot{The identity \sPQJacobi\ motivated the
introduction of the generalized KLT matrix in \PScomb. In the multiparticle pure spinor superfield
framework, where $V_P$ are local superfields satisfying generalized Jacobi identities and $M_Q$ are
Berends-Giele current superfields satisfying shuffle symmetries, we have the relation
$V_{iA} = \sum_B S(A|B)_i M_B$ \Polylogs.  However, the fixed leg $i$  prevents  the Jacobi identities
of $V_{iA}$ being manifest.}

Finally, another consequence of \bracketagain\ is that
\eqn\deccurly{
\sum_{XY=P}\{X,Y\} \sim s_P P\,.
}
This follows immediately from \BMap, by acting with the KLT map on both sides.


\newnewsec\seccontact The contact term map as a Lie co-bracket

A series of studies of string theory correlators and BCJ numerators led to the so-called {\it contact term
map} \genredef\ appearing in the action of the pure spinor BRST operator on local multiparticle superfields \EOMbbs.
This section identifies the contact term map as the Lie co-bracket dual to the $S$-bracket and proves its main
properties as a result of this identification.

\proclaim Definition (Contact term map). The contact term map, $C: \Lie_\cS\rightarrow \Lie_\cS\wedge\Lie_\cS$, is the
dual of the $S$-bracket; i.e. for a Lie monomial, $\Gamma$,
\eqn\basicC{
(P\otimes Q, C(\Gamma)) := (\{P,Q\}, \Gamma).
}
An explicit formula for $C(\Gamma)$ follows from \basicC\ by choosing a basis, for example:
\eqn\defC{
C(\Gamma) = \sum_{P,Q} (\{P,Q\},\Gamma)\; \ell[P]\otimes\ell[Q],
}
where the sum is over a Lyndon basis of $\Lie^*$.\foot{Other explicit formulas follow from choosing
different pairs of dual basis. e.g. we could write a `left-right' version of \defC:
$C(\Gamma) = \sum_{P,Q} (\{P,Q\},\Gamma) \ell[P]\otimes r[Q]$,
which is related to \defC\ by $Q \sim  (-1)^Q \bar Q$ and $\ell[Q] = (-1)^Q r[\bar Q]$.}

Defining $P\wedge  Q := P\otimes Q - Q\otimes P$, the first few examples of the map $C$ are
\eqnn\defCex
$$\eqalignno{
~&C([1,2]) = (k_1\cdot k_2) (1\wedge  2 ),&\defCex\cr
~&C([1,[2,3]]) = (k_2\cdot k_3)\left( [1,2]\wedge  3 + 2\wedge  [1,3]\right)
+ (k_{1}\cdot k_{23}) \left(1 \wedge  [2,3]\right).
}$$

$C$ satisfies the {\it dual Jacobi identity} ($A$ is the swap map, $X\otimes Y\mapsto Y\otimes X$.)
\eqn\dualjacobiidentity{
(C\otimes Id)\circ C - (Id\otimes C)\circ C - (Id\otimes A) \circ (C\otimes 1) \circ C = 0,
}
which is equivalent to the Jacobi identity for the $S$-bracket.
Also, recall the main property,
\eqn\anothertime{
b(\{P,Q\}) = [b(P),b(Q)].
}
This is equivalent to:

\proclaim Lemma. $C$ satisfies \genredef
\eqn\CcircbP{
C(b(P)) = \sum_{P=XY} b(X)\wedge b(Y) .
}\par
\proof
\anothertime\ implies that
\eqn\PQCG{
(P\otimes Q, C(b(R))) = ([b(P),b(Q)], R).
}
The RHS can be expanded by deconcatenation (as in the derivation of \BMap):
\eqn\PQCGdec{
RHS =  \sum_{R=XY} (b(P),X)(b(Q),Y) - (X\leftrightarrow Y).
}
But $b$ is self-adjoint, and so
\eqn\CbPtwo{
(P\otimes Q,C(b(R))) = \sum_{R=XY} (P,b(X))(Q,b(Y)) -  (X\leftrightarrow Y),
}
and this is equivalent to \CcircbP. \CcircbP\ can also be checked using the formula, \defC.\qed

\medskip
 
The rest of this section derives a recursive formula for $C$. 
First define the standard extension of the adjoint representation of $\Lie$ to $\Lie\otimes\Lie$:
\eqnn\otdef
$$\eqalign{
[P,X\otimes Y] &:= [P,X]\otimes Y + X\otimes [P,Y],\cr
[X\otimes Y,Q] &:= [X,Q]\otimes Y + X\otimes [Y,Q].
}$$
This makes $\Lie\otimes\Lie$ into an adjoint representation of $\Lie$. 

\proclaim Lemma (Recursion). For $\Gamma_1,\Gamma_2\in \Lie$, the action of $C$ on $[\Gamma_1,\Gamma_2]$ is given by
\eqn\recurPQ{
C([\Gamma_1,\Gamma_2]) := k_{1}\cdot k_{2}\,  \Gamma_1\wedge \Gamma_2+[C(\Gamma_1), \Gamma_2] + [\Gamma_1,C(\Gamma_2)],
}
where $k_1$ and $k_2$ are the momenta associated to $\Gamma_1$ and $\Gamma_2$. With $C(i) :=0$, \recurPQ\ can be taken as a definition of $C$, as in \genredef.

\proof This is a consequence of the identity, \deltacurlyAB, used in the proof of the defining property of the $S$ bracket, \anothertime. We again use deconcatenation (as in \PQCG, above) to write
\eqn\triviality{
(\{P,Q\},[\Gamma_1,\Gamma_2]) = (\delta \{P,Q\}, \Gamma_1\otimes \Gamma_2) - (1\leftrightarrow 2).
}
Then \deltacurlyAB\ gives
\eqnn\CPQAB
$$\eqalignno{
(P\otimes Q,C([\Gamma_1,\Gamma_2])) &= k_{P}\cdot k_{Q} (P,\Gamma_1)(Q,\Gamma_2) &\CPQAB\cr
&+ \sum_{P=XY} (X,\Gamma_1)(\{Y,Q\},\Gamma_2) -  \sum_{Q=XY} (\{P,X\},\Gamma_1)(Y,\Gamma_2) - (1\leftrightarrow 2).
}$$
But $(\{Y,Q\},\Gamma_2) = (Y\otimes Q,C(\Gamma_2))$ and $(\{P,X\},\Gamma_1) = (P\otimes X, C(\Gamma_1))$, and so \CPQAB is equivalent to \recurPQ.
\qed

The recursive relation, \recurPQ, can be solved to find an explicit formula for $C(\Gamma)$. To see this, write
\eqn\defD{
D(\Gamma) := k_1\cdot k_2\, \Gamma_1\wedge  \Gamma_2,
}
for Lie monomials $\Gamma = [\Gamma_1,\Gamma_2]$. Nesting \recurPQ\ leads to a sum over the edges in the tree $T_\Gamma$. For $I \in \Gamma$ an edge in the tree $T_\Gamma$, let $\Gamma_I$ be the associated Lie monomial. For example, if $\Gamma =[[1,2],[3,4]]$, then $\Gamma_{12} = [1,2]$. Then the solution to the recursion, \recurPQ, is
\eqn\defCexplicit{
C(\Gamma) = \sum_{I\in\Gamma} \Gamma/\Gamma_{I}[D(\Gamma_I)],
}
where $\Gamma/\Gamma_{I}[D(\Gamma_I)]$ denotes the replacement, in $\Gamma$, of $\Gamma_I$ by $D(\Gamma_I)$.

For example, if $\Gamma=[[1,2],[3,4]]$, then $C(\Gamma)$ is
\eqn\Cmapexamplefull{
C(\Gamma) = D(\Gamma) + \Gamma/\Gamma_{12}[D(\Gamma_{12})] + \Gamma/\Gamma_{34}[D(\Gamma_{34})],
}
where
\eqn\Cmapexample{
\Gamma/\Gamma_{34}[D(\Gamma_{34})] = s_{34} [[1,2],3\otimes 4] = s_{34} \left( [[1,2],3]\otimes 4 + 3\otimes [[1,2],4] \right),
}
and
\eqn\Dexample{
D(\Gamma) = k_{12}\cdot k_{34} \, [1,2]\wedge [3,4].
}

\subsecno=1 
\newsubsubsec\pscont The contact-term map and equations of motion

The contact term map plays a role in the equations of motion of the $V_\Gamma$ \refs{\EOMbbs,\Gauge,\genredef}.
Associated to the $V_\Gamma$ are SYM Berends-Giele currents, $M_P = V(b(P))$, where $V(\Gamma) := V_\Gamma$.
For example,
\eqn\Mexs{
M_1 = V_1, \quad M_{12} = {V_{[1,2]}\over s_{12}},\quad
M_{123} = {V_{[[1,2],3]}\over s_{12}s_{123}} + {V_{[1,[2,3]]}\over s_{23}s_{123}}\,.
}
The equation of motion of $M_{P}$ under the action of the pure spinor BRST charge $Q$ is given by \Gauge
\eqn\QMP{
QM_P = \sum_{XY=P}M_XM_Y\,.
}
Whereas the equation of motion for the local superfields $V_\Gamma$ can be written as
\eqn\QVG{
QV_\Gamma = \half (V\otimes V, C(\Gamma)) = \half\sum_{P,Q} V_{\ell(P)}V_{\ell(Q)} (P\otimes Q,C(\Gamma)),
}
where the sum is over a basis. For example,
\eqnn\VPQex
$$\eqalignno{
QV_{[1,2]}&=(k_1\cdot k_2) V_1V_2,&\VPQex\cr
QV_{[[1,2],3]} &= (k_1\cdot k_2)(V_{[1,3]}V_2 + V_1V_{[2,3]}) + (k_{12}\cdot k_3)V_{[1,2]}V_3,\cr
QV_{[1,[2,3]]} &= (k_2\cdot k_3)\big(V_{[1,2]}V_3 + V_2V_{[1,3]}\big) + (k_1\cdot
k_{23})V_1V_{[2,3]}\,,
}$$
and it can be checked that these imply \QMP\ for $P=123$. In fact, given \CcircbP, equation
\QVG\ implies equation \QMP, as explained in \genredef. This was previously only known from explicit
calculations at low multiplicities.

\newnewsec\Atheorysec BCJ numerators

In the previous sections we used Berends-Giele currents to recast and prove relations for our Lie polynomial version of
biadjoint scalar theory.  When dressed with BCJ numerators these results apply directly to other coloured theories.  In this
section, we first argue that any coloured theory that admits the Berends-Giele framework can be understood in this way. There
are clear distinctions between the behaviour of numerators for on-shell amplitudes, versus those for the partially off-shell
Berends-Giele currents; for a given Berends-Giele description we will see that numerators are unique, but the
Berends-Giele description itself has a gauge freedom as exploited in \Gauge.  We then discuss explicit examples of numerators $N^{\rm
theory}_\Gamma$ for different theories.

\newsubsec\BCJcoloured BCJ numerators for coloured theories

Many gauge theories have perturbation expansions that can be expanded in colour factors of the form \colourfactor; this is the
case for any gauge theory whose Lagrangian is second order and single trace in the Lie algebra, with interaction terms formed
out of total Lie bracketings. For such a theory, there is an iteration for $\Lie$-valued fields, analogous to the one for
biadjoint scalar in \S\BGsec\ and for Yang-Mills in appendix A.  For a particular gauge choice, this iteration generates colour
ordered Berends-Giele currents, $B^{\rm theory}(P)$, with the partial amplitudes of the theory given by $\cA^{\rm
theory}(Pn)=\lim_{s_P\rightarrow 0}s_P B^{\rm theory}(P)\cdot \epsilon_n$, where $\epsilon_n$ is the polarization of the $n$th
particle. These partial amplitudes trivially satisfy shuffle relations \BGsym\ as a consequence of the BG currents being Lie
polynomials.  The amplitudes $\cA(Pn)$ are invariant under field redefinitions and gauge transformations, but the Berends-Giele
currents $B^{\rm theory}(P)$ are not invariant. For a coloured theory in the double copy, we now consider how the currents $B^{\rm
theory}(P)$ are related to $b(P)$.

A BCJ numerator associates every Lie monomial $\Gamma$ with a function $N_\Gamma ^{\rm
theory}\in {\cal K^{\rm theory}}$.  Here $\cK^{\rm theory}$ is the space of functions of the kinematic data. For example, for the nonlinear sigma model, ${\cal K}^{\rm NLSM}$ are  functions of the Mandelstam invariants. For Yang Mills, ${\cal K}^{\rm YM}$ consists of  functions that are also multilinear in the polarisation vectors of the gluons. As observed in \Frost, the statement of colour kinematics duality is that $\Gamma \mapsto N_\Gamma^{\rm theory}$ defines a homomorphism 
\eqn\BCJnum{
N^{\rm theory}:\Lie\rightarrow {\cal K^{\rm theory}}\,.
}
Instead of considering BCJ numerators for {\it amplitudes}, we can look for maps, $\tilde N^{\rm theory}$, which relate BG currents to $b(P)$:
\eqn\BPbP{
B^{\rm theory}(P)=\tilde N^{\rm theory}(b(P))\,.
}
In other words, we regard $\tilde N_\Gamma$ as (off-shell) BCJ numerators for the BG currents of the theory. The numerators
$\tilde N_\Gamma$ may have free indices that $N_\Gamma$ does not have (e.g. for YM, $\tilde N_\Gamma$ has a free gluon polarization
index). This almost trivial change of perspective has an interesting consequence.

Dropping the superscript `theory' from now on, fix some gauge theory, and, assuming some gauge choices, let $B(P)$ be the BG
currents of the theory in some gauge. The results of \S\secklt\ show that we can invert $b(P)$ using \bInvs\ to write
\eqn\discussN{
\tilde N_\Gamma :=  B(\{\Gamma\})\, .
}
Here $\{\Gamma\}$ is only well-defined as an element of $\Lie^*$, but $B(\{\Gamma\})$ is well defined since, as we explain
above, $B^{\rm theory}(R\shuffle S)=0$ for $R,S\neq \emptyset$. Moreover, $\Gamma \mapsto \tilde N_\Gamma$ is a homomorphism,
because the $S$-bracket is Lie. The key point is that, given some set of $B(P)$, the $\tilde N_\Gamma$ defined by \discussN\
are unique.

It might be helpful to describe \BPbP\ and \discussN\ more explicitly in a basis. Using \curlybasisexpand,
\eqn\numerator{
\tilde N_{\ell[P]}=  B(\ell\{P\})=\sum_{Q\in Basis}S^\ell(P|Q) B(Q)\, ,
}
where $S^\ell(P|Q)$ is the generalized KLT matrix. If we multiply by $s_P$, contract free indices with
the $n$th particle, and take the limit $s_P\rightarrow 0$, \numerator\ becomes a well known formula
for amplitude BCJ numerators, given in \DuNLSM (see also the discussion in \S5 of \NLSM). Then \BPbP\ reads
\eqn\discussNbP{
N(b(P))=\sum_{Q\in Basis} N_{\ell[Q]} \, b(P|Q)= \sum_{Q,R\in Basis} S^\ell(Q|R) \, B(R)\, b(P|Q)=B(P)\, .
}
Thus the existence of `off-shell' numerators is generic and unique, given a choice of
Berends-Giele currents.

This is in sharp contradistinction with the {\it on-shell} BCJ numerators for amplitudes.
On-shell BCJ numerators are subject to a gauge freedom spanned by numerators of the form  
\eqn\numgauge{
N^{\rm gauge}_\Gamma =  \sum_{R,S} C_{R,S} (\{R,S\},\Gamma),
}
for some arbitrary kinematic functions $C_{R,S} \in \cK$. These contribute the following to the BG currents: 
\eqn\OnShgauge{
N_\Gamma^{\rm gauge} ( b(P))=  \sum_{R,S} C_{R,S}  b(P|\{R,S\})=   \sum_{R,S} C_{R,S} (P, [b(R),b(S)])\, ,
}
This contribution vanishes on-shell, because the RHS \OnShgauge\ has no $1/s_P$ pole, and so
vanishes when multiplied by $s_P$, in the $s_P\rightarrow 0$ limit.  It was this gauge freedom
in a different guise that led to the original discovery of the BCJ relations in \BCJ, where it
was argued that there are $(n-2)!-(n-3)!$ independent pure gauge numerators of this form. However,
\OnShgauge\ shows that these no longer vanish off-shell. The off-shell numerators are therefore not
subject to the freedom, \numgauge, and are unique once a choice of Berends Giele formulation has been made.

BCJ numerators for amplitudes, $N_\Gamma$, are said to be {\it local} if they contain no poles in
the Mandelstam variables.\foot{For a coloured theory, BCJ numerators can be obtained when the
fundamental BCJ identities are satisfied by inverting on an $(n-3)!$ basis, but
this will lead to spurious poles in general.} Local BCJ numerators are known for both the nonlinear sigma model (NLSM),
\refs{\DuNLSM,\NLSM}
and for (super-)Yang-Mills \refs{\MafraKJ,\BGBCJ,\genredef,\dufufeng,\ET,\Mizcop} (see also \BGFthree\ for theories with
deformations $\ap F^3$ and $\ap^2 F^4$) see \JJreview\ for a review.

If the off-shell numerators of a theory are local, they restrict to give local on-shell numerators
for the amplitudes of the theory.\foot{An arbitrary set of  local on-shell BCJ numerators for the
theory may not be the restriction of the off-shell numerators, because of the gauge freedom, \numgauge.}
But, for a generic theory  there is no reason to expect that the unique {\it off-shell} numerators obtained
from \discussN\ will be  {local}.  Indeed,  an obstruction to locality  for Yang-Mills Berends-Giele
currents in Lorenz gauge is identified in \refs{\BGBCJ,\genredef,\Gauge}.
This led the authors to the introduction of `BCJ gauge' in which local numerators are obtained.

 If a theory does not admit any set of local off-shell numerators, this would prevent it from participating in a KLT relation:
\eqn\double{
M=\sum_{P,Q\in Basis} S^\ell(P|Q)B(P)\tilde B(Q)\,
}
where $\tilde B(Q)$ might be another coloured theory.  Since  $S^\ell(P|Q)$ cancels one copy of $b(P)$ in
the product, locality of the numerators implies that the singularities of $M$ arise only from one copy of
$b(P)$. In other words, $M$ has singularities consistent with amplitude factorization. Without locality
of the numerators, the singularity and factorisation structure of $M$ would not be consistent, and so
$M$ would not come from a local field theory. The existence of local numerators is therefore what
makes the theories in the BCJ web special.  

If the off-shell numerators, \discussN, of a gauge theory are local, this implies that the theory
satisfies the on-shell BCJ relations. Indeed,
 \eqn\whyitis{
 B(\{P,Q\})=N(b(\{P,Q\})).
 }
If $N$ is local, \whyitis\ will have no $1/s_{PQ}$ pole, since $b(\{P,Q\})$ has no $1/s_{PQ}$ pole.
Then the partial amplitudes associated to $B(P)$ satisfy
\eqn\whathappens{
A^{\rm theory}(\{P,Q\},n)= \lim_{s_{PQ}\rightarrow 0} s_{PQ} B^{\rm theory} (\{P,Q\})\cdot\epsilon_n=0\,,
}
which are the fundamental BCJ relations. Given this, it is reasonable conjecture the converse: that if a theory's partial
amplitudes satisfy the BCJ relations, then there exists a field redefinition and gauge fixing of its Berends-Giele recursion so
that it has {\it local} numerators. As discussed in \BGBCJ, for (super)Yang-Mills the locality
of the numerators following from Berends-Giele currents as
in \discussN\
is achieved in the so-called {\it BCJ gauge} \Gauge. This gauge is characterized by multiparticle fields labelled by
planar binary trees satisfying generalized Jacobi identities and can be obtained via a standard finite gauge transformation
of the gauge (super)fields \refs{\Gauge,\genredef}.

The following subsections review particular examples.

\newsubsec\Atheorybiadjointsec Biadjoint scalar

For the biadjoint scalar theory,\foot{This was an early example of the benefits of using combinatorial structures; biadjoint
scalar tree amplitudes $m(Pn|Qn)$ were obtained in \DPellis\ from planar binary trees. As shown in \FTlimit, the solution to the
biadjoint scalar field equations yields a recursion for Berends-Giele currents $b(P|Q)$ depending on two color
orderings from which the amplitudes are obtained via the Berends-Giele amplitude formula. In \PScomb, these double fields were
reformulated as a map on the binary tree expansion \BMap, given by \biadjointbg.} the numerators take values in $\cK^{\rm
Biadjoint} =W$ i.e., in words $Q$ representing colour orderings for the second Lie algebra:
\eqn\biajNum{
N(Q)^{\rm Biadjoint}_\Gamma=(Q,\Gamma)\, .
}
If $\Gamma_1+\Gamma_2+\Gamma_3=0$ follows from the  Jacobi identity, then
$N_{\Gamma_1} + N_{\Gamma_2} + N_{\Gamma_3} =(Q,\Gamma_1+\Gamma_2+\Gamma_3)=0$ follows immediately.
This yields
$
b(P|Q)=(Q,b(P))
$
as defined earlier.

Following \DPellis, it was pointed out in \FTlimit\ that this Berends-Giele multiparticle field $b(iP|iQ)$ gives rise to an efficient
algorithm to compute the inverse of the KLT matrix $S(P|Q)_i$, but no direct algebraic proof was given for this
statement \PScomb\foot{The statement that the KLT matrix is the inverse to the ``biadjoint amplitudes'' was argued on general grounds
using intersection theory by Mizera \mizera. The question in \PScomb\ was whether the two matrices $S(P|Q)_i$ and $b(P|Q)$ defined by the precise recursive formulas given in \NLSM\ and \FTlimit\ could be shown to be mutually inverse directly.}.
This has now been proven in section~\secklt.

The biadjoint amplitudes $m(Pn|Qn)$
are then given by projecting the binary tree expansion $b(Q)$ of \BGpbt\ against $P$ \PScomb,
\eqn\biadjoint{
m(Pn|Qn) = \lim_{s_P\rightarrow 0} s_P(P, b(Q))) = \lim_{s_P\rightarrow 0}  s_P\,b(P|Q)\,.
}
The Kleiss-Kuijf relations follow from Ree's theorem \reestheorem\
as $b(R\shuffle S|Q) = 0$ because $b(Q)$ is a Lie polynomial, while the BCJ relations follow
from the $\{,\}$-bracket as discussed in section~\BCJsec.

Example of biadjoint amplitudes are obtained from $b(123)$ and $b(1234)$:
\eqnn\exbi
$$\eqalignno{
m(1234|1234) &=
{\langle 123,[[1,2],3]\rangle\over s_{12}}
+ {\langle 123,[1,[2,3]]\rangle\over s_{23}} ={1\over s_{12}}+{1\over s_{23}}, &\exbi\cr
m(12345|14235) &=
{\langle 1234, [ [ [ 1 , 4 ] , 2 ] , 3 ] \rangle\over s_{14} s_{124}}
+  {\langle 1234, [ [ 1 , [ 4 , 2 ] ] , 3 ] \rangle\over s_{124} s_{24}}
+  {\langle 1234, [ [ 1 , 4 ] , [ 2 , 3 ] ] \rangle\over s_{14}  s_{23}}\cr
&+  {\langle 1234, [ 1 , [ [ 4 , 2 ] , 3 ] ] \rangle\over s_{24} s_{234}}
+  {\langle 1234, [ 1 , [ 4 , [ 2 , 3 ] ] ] \rangle\over s_{234} s_{23}}= - {1\over s_{23}s_{234}}\,,
}$$
where the expansion of $b(123)$ and $b(1234)$ can be found in \bexampone.

\newsubsec\AtheoryNLSMsec NLSM

NLSM amplitudes can be studied by BG recursion \BGTrnka, as above for biadjoint scalar theory. Although we do not prove it
here, experimental evidence suggested the following formula for (off-shell) BCJ numerators for NLSM:
\eqn\AtheoryNLSM{
N^{\rm NLSM}_\Gamma := \sum_{P\in Basis} (P,\Gamma)S^\ell(P|P), \quad\hbox{for $\len{\Gamma}=n$}
}
where $S^\ell(P|Q)$ is the generalized KLT matrix. It is clear that $\Gamma \mapsto N^{\rm NLSM}_\Gamma$ is a homomorphism. In particular, if $\Gamma_1+\Gamma_2+\Gamma_3=0$ is a triple of Lie monomials that vanishes, then the corresponding numerators satisfy the BCJ numerator
identity $N_{\Gamma_1} + N_{\Gamma_2} + N_{\Gamma_3} =0$. We refer to \FrostDPhil\ for an
on-shell proof of \AtheoryNLSM\ using CHY methods. 

The NLSM amplitudes are computed
from the correspondence \BGpbt,
\eqn\NLSMamp{
A^{\rm NLSM}(Pn) = \lim_{s_P\rightarrow 0}N^{\rm NLSM}\big(s_P b(P)\big)\,.
}
For example, at 4-points the numerators are
\eqn\nlsmgammast{\eqalign{
N^{\rm NLSM} ([[1,2],3]) &= s_{12}(s_{23}+s_{13}),\cr
N^{\rm NLSM}([1,[2,3]]) &= s_{12}(s_{23}+s_{13}) - s_{13}(s_{23}+s_{12}) = s_{23}(s_{12}-s_{13}),\cr
N^{\rm NLSM}([[1,3],2]) & = s_{13}(s_{23}+s_{12}).
}}
These satisfy
\eqn\satisfy{
N^{\rm NLSM} ([[1,2],3]) - N^{\rm NLSM}([1,[2,3]]) - N^{\rm NLSM}([[1,3],2]) = 0.
}
Acting on $b(123)$, we get
\eqn\fourN{
s_{123}N^{\rm NLSM}( b(123)) = s_{12}+s_{23}\,,
}
which is the 4 point partial amplitude, $A^{\rm NLSM}(1234)$. From the existence of the map \AtheoryNLSM, and the discussions in \S\secbcj, it follows that the KK and BCJ relations are automatically satisfied by the NLSM amplitudes, as was proved in \sigmaDu. In \NLSM, master BCJ numerators with fixed legs $1$ and $n$
of the NLSM amplitudes were observed in examples to be $N_{1|P|n}=(-1)^{n/2}S(P|P)_1$ for even $n$.
\AtheoryNLSM\ has this as a special case.

\newsubsec\QBRSTCsec Super-Yang--Mills

String theory OPEs (or supersymmetric BG recursion) can be used to recursively compute
local SYM multiparticle superfields $\{A^\Gamma_\a,A^\mu_\Gamma,W^\a_\Gamma,F^{\mu\nu}_\Gamma\}$, $\mu,\nu=1,\ldots ,10$, $\alpha=1,\ldots, 16$, in the BCJ
gauge which are labelled by Lie monomials $\Gamma \in \Lie$ \refs{\EOMbbs,\Gauge,\genredef}. These can be taken to be the SYM numerators, i.e.\ $N_{\mu\Gamma}^{SYM}=A_{\mu\Gamma}$ etc..
As demonstrated in \refs{\Gauge,\genredef}, the words labelling these superfields
satisfy the same generalized Jacobi identities associated with the corresponding Lie monomial $\Gamma$. For example,
\eqn\Amex{
A^\mu_{[[1,2],[3,4]]} = A^\mu_{[[[1,2],3],4]} -A^\mu_{[[[1,2],4],3]}.
}
This leads to a proposal for local BCJ-satisfying numerators $N_\mu^{\rm SYM}$ from which
SYM tree amplitudes arise from
\eqn\SYMbmap{
\cA^{\rm SYM}(Pn) =A^\mu_n\lim_{s_P\rightarrow 0} s_P N_\mu^{\rm SYM} \big( b(P)\big),
}
where $A^\mu_n$ is the polarization vector of the $n$th particle while the action of $N_\mu^{\rm SYM}$
on the Lie polynomials $\Gamma=[\Gamma_1,\Gamma_2]$ in \BMap\ 
is given by
\eqn\AYMmapPrin{
A^\mu_nN_{\mu\Gamma}^{\rm SYM}
:=
 A_{n\mu} A^\mu_{[\Gamma_1,\Gamma_2]}\, ,
}
in terms of the $\t=0$ component of the local multiparticle superfield $A^\mu_{[\Gamma_1,\Gamma_2]}$ \refs{\EOMbbs,\genredef}.
This representation {\it manifestly} satisfies the BCJ identities. For example,
the five-point color-ordered amplitudes in the
Kleiss--Kuijf basis following from the maps \SYMbmap\ and \AYMmapPrin\ are given by
\eqnn\ampsfive
$$
\cA(12345) =
\Big({A^\mu_{[ [ [ 1 , 2 ] , 3 ] , 4 ]} \over s_{12} s_{45}}
+  {A^\mu_{[ 1 , [ [ 2 , 3 ] , 4 ] ]} \over s_{23} s_{51}}
+  {A^\mu_{[ [ 1 , 2 ] , [ 3 , 4 ] ]} \over s_{12} s_{34}}
+  {A^\mu_{[ [ 1 , [ 2 , 3 ] ] , 4 ]} \over s_{45} s_{23}}
+  {A^\mu_{[ 1 , [ 2 , [ 3 , 4 ] ] ]} \over s_{51} s_{34}}\Big) A_{5\mu}
$$
together with the other 5 permutations of $2,3,4$.
It is straightforward to check that all BCJ numerator identities are manifestly satisfied.
For example, comparing the
above parametrization with the one in \BCJ\ leads to
\eqn\nnums{
n_3 = A^\mu_{[ [ 1 , 2 ] , [ 3 , 4 ] ]}A^\mu_5,\qquad
n_5 = A^\mu_{[ 1 , [ 2 , [ 3 , 4 ] ] ]}A^\mu_5,\qquad
n_8 = A^\mu_{[ [ 1 , [ 4 , 3 ] ] , 2 ]}A^\mu_5\,,
}
from which the identity $n_3-n_5+n_8=0$ can easily be verified.

\newsubsubsec\nonlocN Non-local vs local BCJ numerators from Berends-Giele in the Lorenz vs BCJ gauge

As an example of the definition of BCJ-satisfying numerators \discussN\ and its (non)locality properties, consider
the Berends-Giele current of standard Yang-Mills theory \BGpaper\ for the color-ordered five-point amplitude.
As shown in \BGBCJ, the standard Berends-Giele current $J^\mu(1234)$ of \BGpaper\ can be written in terms of multiparticle 
fields $\hat A^\mu_{[\Gamma_1,\Gamma_2]}$ in the
{\it Lorenz gauge}
\eqn\stdBG{
s_{1234}J^\mu(1234) = {\hat A^\mu_{[ [ [ 1 , 2 ] , 3 ] , 4 ]} \over s_{12} s_{45}}
+  {\hat A^\mu_{[ 1 , [ [ 2 , 3 ] , 4 ] ]} \over s_{23} s_{51}}
+  {\hat A^\mu_{[ [ 1 , 2 ] , [ 3 , 4 ] ]} \over s_{12} s_{34}}
+  {\hat A^\mu_{[ [ 1 , [ 2 , 3 ] ] , 4 ]} \over s_{45} s_{23}}
+  {\hat A^\mu_{[ 1 , [ 2 , [ 3 , 4 ] ] ]} \over s_{51} s_{34}}\,,
}
where $\hat A^\mu_{[\Gamma_1,\Gamma_2]}$ are the recursive multiparticle vector potential
\eqn\AmLrecurs{
\hat{A}_\mu^{[\Gamma_1,\Gamma_2]}=-\half\big[\hat A^{\Gamma_1}_\mu(k^{\Gamma_1}\cdot \hat A^{\Gamma_2}) +
\hat A^{\Gamma_1}_\nu \hat F_{\mu\nu}^{\Gamma_2}
-(\Gamma_1\leftrightarrow \Gamma_2)\big]
}
and $F_{\mu\nu}^{\Gamma_2}$ are the local multiparticle field-strengths defined similarly \refs{\Gauge,\genredef}, see for example \S4.3 of \genredef\ for the difference between $A_\Gamma^\mu$ and $\hat A_\Gamma^\mu$.

Using the prescription \discussN\ to obtain BCJ-satisfying numerators one gets, for example
\eqn\Numcur{
\eqalign{
N^\mu_{[[1,2],[3,4]]} &= J^\mu(\{\{1,2\},\{3,4\}\})\,,\cr
N^\mu_{[1,[2,[3,4]]]} &= J^\mu(\{1,\{2,\{3,4\}\}\})\,,
}\qquad
\eqalign{
N^\mu_{[[1,[4,3]],2]} &= J^\mu(\{\{1,\{4,3\}\},2\})\cr
\phantom{m}&\cr
}
}
leading to
\eqnn\NumExa
$$\eqalignno{
N^\mu_{[[1,2],[3,4]]} &= s_{12}s_{34}\big(
        J^\mu(1234) s_{23}
       - J^\mu(1243) s_{24}
       - J^\mu(2134) s_{13}
       + J^\mu(2143) s_{14}
      \big)\,,\qquad{} &\NumExa\cr
N^\mu_{[1,[2,[3,4]]]} &=
       s_{23}s_{34}\big( J^\mu(1234) s_{12}
       - J^\mu(1324) s_{13}
       - J^\mu(1342) s_{13}
       + J^\mu(1432) s_{14}\big)\cr
       &+ s_{24}s_{34}\big(J^\mu(1423) s_{14}
       + J^\mu(1432) s_{14}
       - J^\mu(1243) s_{12}
       - J^\mu(1342) s_{13}\big)\,,\cr
N^\mu_{[[1,[4,3]],2]} &=
       s_{13}s_{34}\big( J^\mu(1432) s_{23}
       - J^\mu(1342) s_{24}
       - J^\mu(4312) s_{12}
       + J^\mu(4132) s_{23}\big)\cr
       &+s_{14}s_{34}\big( J^\mu(1432) s_{23}
       - J^\mu(3142) s_{24}
       - J^\mu(1342) s_{24}
       + J^\mu(3412) s_{12}\big)\,.
}$$
It is easy to see that the BCJ numerator relation $n_3-n_5+n_8=0$ is satisfied by the above representations
as $N^\mu_{[[1,2],[3,4]]} - N^\mu_{[1,[2,[3,4]]]} + N^\mu_{[[1,[4,3]],2]} =0$
due to the shuffle symmetry $J^\mu(R\shuffle S)=0$ (note that the numerators follow from contraction
with the polarization $A^\mu_5$, e.g. $n_3=N^\mu_{[[1,2],[3,4]]} A^\mu_5$).
For example, the terms proportional to $s_{14}s_{24}s_{34}$ are given
by
\eqn\Jzero{
          - J^\mu(1342)
          - J^\mu(1423)
          - J^\mu(1432)
          - J^\mu(3142) = - J^\mu(3\shuffle 142) = 0\,.
}
All the other BCJ numerator relations can be similarly verified.

It can be checked by explicitly computing the BG currents in Lorenz gauge that the numerators \NumExa\ are {\it not} local.
But when the Berends-Giele
currents are computed in the BCJ gauge,
the right-hand side of \NumExa\ do become local expressions.
Indeed, using $A^\mu_{[\Gamma_1,\Gamma_2]}$ in the BCJ gauge
(instead of $\hat A^\mu_{[\Gamma_1,\Gamma_2]}$ in \stdBG),
the linear combinations in \NumExa\ yield:
\eqn\NumBCJ{
\eqalign{
N^\mu_{[[1,2],[3,4]]} &= A^\mu_{[[1,2],[3,4]]}\cr
N^\mu_{[1,[2,[3,4]]]} &= A^\mu_{[1,[2,[3,4]]]}
}\qquad
\eqalign{
N^\mu_{[[1,[4,3]],2]} &= A^\mu_{[[1,[4,3]],2]}\cr
\phantom{m}&\cr
}
}
To see this note that in the BCJ gauge $A^\mu_{[\Gamma_1,\Gamma_2]}$ satisfies the
same relations as the Lie monomial $[\Gamma_1,\Gamma_2]$ so \NumBCJ\ follow from the property $\Gamma=b(\{\Gamma\})$
proved in \curlylabla.

\newsubsec\Atheorybiadjointsec Z-theory and the open superstring

In order to upgrade the discussion in the previous subsection to the open superstring with $\ap$ corrections
we will exploit the non-abelian Z-theory method to evaluate $\ap$ expansions of open string disk integrals.
Recall that the string disk integrals are computed via the Berends-Giele method as
\eqn\Zints{
Z(P,n|Q,n) = \lim_{s_P\rightarrow 0}s_P b^\ap(P|Q)
}
where the Berends-Giele currents are computed using the equations of motion of the non-abelian Z-theory \BGap.
One can promote the setup of \BGap\ to the theory of free Lie algebras by assuming the existence
of $\ap$ corrections to the Catalan expansion \BMap\ and defining
\eqn\phiapdef{
b^\ap(P|Q) = (b^\ap(P), Q)
}
which together with \Zints\ implies that Z-theory admits a free Lie algebra representation.
Using the explicit expressions of $b^\ap(P|Q)$ up to $\ap^7$ order from \BGap\ 
one can show that they indeed admit a Lie-polynomial form,
\eqnn\BMapap
$$\eqalignno{
s_P b^\ap(P) &= \sum_{XY=P}[b^\ap(X),b^\ap(Y)] &\BMapap\cr
&+ \ap^2\zeta_2\sum_{XYZ=P}k_X\cdot k_Y[b^\ap(X),[b^\ap(Z),b^\ap(Y)]]\cr
&- \ap^2\zeta_2\sum_{XYZ=P}k_Y\cdot k_Z[[b^\ap(X),b^\ap(Y)],b^\ap(Z)]\cr
&+ \ap^2\zeta_2\sum_{XYZW=P}[[b^\ap(X),b^\ap(Y)],[b^\ap(W),b^\ap(Z)]]\cr
&- \ap^2\zeta_2\sum_{XYZW=P}[[b^\ap(X),b^\ap(Z)],[b^\ap(W),b^\ap(Y)]] +{\cal O}(\ap^3)
}$$
and so on.

It is important to emphasize that the symmetries of the ``domain'' $P$ and integrand $Q$ are different, in
particular $b^\ap(P|Q)\neq b^\ap(Q|P)$.
The integrand $Q$ satisfies shuffle symmetry as $b^\ap(P|R\shuffle S) = \langle b^\ap(P), R\shuffle S\rangle = 0$
because
$b^\ap(P)$ is a Lie polynomial. The shuffle symmetries of the domain $P$ are twisted by the monodromies
of the disk integrals as  explained in \BGap.

Finally, our proposal for obtaining bosonic\foot{Fermionic amplitudes require corrections following from
a Dirac term $\sim k^m_P W_P \gamma^m W_n$ \BGBCJ.} open superstring disk amplitudes including $\ap$ corrections
is to replace the recursion $b(P)$ from \BMap\ by
its $\ap$-corrected $b^\ap(P)$ from \BMapap,
\eqn\disk{
\cA^{\rm string}(P,n) =
\lim_{s_P\rightarrow 0} s_P  N_\mu^{\rm SYM}(  b^\ap(P) ) A_n^\mu\,.
}
This proposal has been explicitly verified for the bosonic components of the disk amplitudes up to seven points and up to
$\ap^3$ \FORM.
While the SYM amplitudes in the field-theory limit $\ap\to0$ are correctly obtained using the analogous 
map $N_\mu^{\rm SYM}$ in the Lorenz gauge, the higher $\ap$ orders of the string disk amplitude \disk\ require the BCJ gauge
as in the map \AYMmapPrin.

\newnewsec\concsec Conclusions

We have seen that there is much nontrivial structure in the amplitudes of coloured  theories that can be formulated in the language of Lie polynomials and their dual.  These  impact quite generally on the study of  tree-level
scattering amplitudes for the theories encompassed by the double copy \JJreview.  Important outstanding questions concern the locality of numerators and the existence and role of the  kinematic algebras that should underpin them.  
We have seen that off-shell Berends-Giele numerators are unique up to the choices of Berends-Giele recursion itself. Since numerators are unique within a Berends-Giele framework, the locality of the corresponding numerators is a determined question.  However, this question of their locality in the off-shell regime is subject to the  choices involved in field redefinition and gauge; we have seen that they are not local for SYM  in Lorenz gauge, but that a BCJ gauge can be found. 
A reasonable conjecture is that a Berends-Giele framework can be found with local numerators for any such theory that factorizes and whose amplitudes satisfy the BCJ relations.

An ongoing topic of research is the identification of the kinematic algebra for a given theory, see  \JJreview\ and references therein. For a given theory, take some set of BG current, $B(P)$. Then these $B(P)$, then we can implicitly define a kinematic bracket $\{,\}_\cK$ on the associated  off-shell numerators $N_{\Gamma_1}$ and $N_{\Gamma_2}$ by imposing the relation
\eqn\kin{
N_{[\Gamma_1,\Gamma_2]} = B(\{[\Gamma_1,\Gamma_2]\}) = \{ B(\{\Gamma_1\}),B(\{\Gamma_2\})\}_\cK = \{N_{\Gamma_1},N_{\Gamma_2}\}_\cK,
}
where $\{~,~\}_\cK$ is the antisymmetric operation, implicitly defined by \kin\ on the image of $N$. It is a Lie bracket on the image of $N$, because $N$ is a homomorphism. Imposing linearity in Mandelstam variables, the definition \kin\ implies an off-shell BCJ-like relation,
\eqn\BCJlike{
B(\{P,Q\}) = \{B(P),B(Q)\}_\cK,
}
where $\{~,~\}_\cK$ is extended to act on the $B(P)$ as
\eqn\kinematicalgeba{
\{B(P),B(Q)\}_\cK := \sum \frac{(P,\Gamma_1)}{s_{\Gamma_1}} \frac{(Q,\Gamma_2)}{s_{\Gamma_2}} \{N_{\Gamma_1},N_{\Gamma_2}\}_\cK.
}
(See \gangchen\ for an off-shell BCJ relation for NLSM, but which is not obviously of the form \BCJlike.) It is highly nontrivial to find a BG formulation that gives local numerators for NLSM and YM. But if a particular choice of gauge is known to give local off-shell numerators, then \kin\ expresses these local numerators as nested bracketings with respect to $\{~,~\}_\cK$. In this situation, it would therefore be reasonable to call $\{~,~\}_\cK$ the ``kinematic algebra'' of the theory; a clear goal in the subject is to find an intrinsic definition of such a bracket on $\cK$.  However, even given a local Lie bracket $\{~,~\}_\cK$ on the  $N_\Gamma$, there is no guarantee that this bracket admits a consistent extension to a Lie bracket
\eqn\kindream{
\{~,~\}_\cK : \cK_P \times \cK_Q \rightarrow \cK_{PQ},
}
on the whole of the kinematic spaces $\cK_P$.

There are many other directions to explore.
Intermediate steps in the calculations of $\ap$ expansion method from \drinfeld\ can be
described by the same combinatorics as the contact terms in the equation of motion for $QV_P$ \OSpriv,
therefore they should profit from framing the results in terms of the contact term map acting 
on Lie polynomials.
Given that the Drinfeld associator itself
is a {\it Lie series} this connection promises to reveal more synergies between mathematical ideas
and scattering amplitudes. Similarly, the Berends-Giele formulation of the non-abelian Z-theory
to compute $\ap$ corrections to string disk integrals \BGap\ induces $\ap$ corrections to the
recursion \BMap\ of planar binary trees as seen in \BMapap. It is natural to ask whether the recursion \BMapap\
itself is generated recursively from purely combinatorial methods. The appearance of factors of $k_X\cdot k_Y$
suggest that the $S$-bracket may play a role. The combinatorial understanding of the Drinfeld method discussed in \andre\
is an encouraging development in this direction.

Given that the field-theory KLT matrix $S^\ell(P|Q)_i$ is the inverse of the biadjoint Berends-Giele double current $b(iR|iS)$
as in \invKLT, it
is natural to look for the inverse of the $\ap$-corrected double current $b^\ap(P|Q)$ from \BGap\ to obtain $\ap$
corrections to the KLT matrix. Note that these corrections will not coincide
with those in the string theory KLT matrix since odd and multiple zeta values are absent
from the latter (see e.g. \refs{\MomKer,\mizKLT}) but must be present in the former.

Similarly, the structures studied in this paper can be realized in the context of the CHY formulae, ambitwistor strings and the
geometry of ${\cal M}_{0,n}$.  Much is already well known, for example \refs{\CBCJ,\CHYKLT,\Frost}, but more appears in
\FrostDPhil.  A particular challenge is to better understand numerators from this perspective.

Again another challenge is to take these insights to higher loops. Tree level colour factors are labelled by Lie monomials, and
partial tree amplitudes are labelled by permutations. This is just the leading order avatar of the more general story, at
arbitrary orders in the perturbation series, in which colour factors are associated to ribbon graphs, and partial amplitudes are
labelled by marked surfaces with boundary (possibly with punctures and nontrivial genus). The results in the present paper are
essentially all derived from the Jacobi identity satisfied by Lie monomials. More generally, colour factors labelled by
ribbon graphs at higher order satisfy analogous identities; again more appears in \FrostDPhil.

The maps and techniques discussed in this paper show that
free Lie algebras permeate the theory of scattering amplitudes. They
also raise the expectations that many more elegant results are yet to be discovered.

\bigskip
\noindent{\bf Acknowledgements:} CRM thanks Oliver Schlotterer for collaboration on closely related topics and for comments on
the draft.
CRM is supported by a University Research Fellowship from the Royal Society.  LJM is supported in part by the STFC grant ST/T000864/1.  HF is supported by ERC grant GALOP ID: 724638.

\appendix{A}{ Berends-Giele recursion for Yang--Mills in the free Lie algebra}
\applab\BGreview

\noindent Here we repeat the discussion of  section \BGsec\ replacing the biadjoint scalar by Yang-Mills.
Consider pure Yang--Mills theory in $d$-dimensions with the Lagrangian
\eqn\YMlag{
{\cal L}_{YM} = {1\over 4}\tr(\bF_{\mu\nu}\bF^{\mu\nu}),\qquad \bF_{\mu\nu} := -[\nabla_\mu,
\nabla_\nu]\,.
}
The trace is over the generators $t^a$ of a Lie algebra, $\cg$,
the covariant derivative is given by $\nabla_\mu = \p_\mu - \bA_\mu$
and $\bA_\mu = \bA_\mu^a t^a$ is the gluon potential.
In the Lorenz gauge, $\p_\mu \bA^\mu = 0$, the field equation $[\nabla_\mu, \bF^{\mu\nu}]
= 0$ becomes
\eqn\EOMYM{
\Box \bA^\nu(x) = [\bA_\mu(x),\p^\mu\bA^\nu(x)] + [\bA_\mu(x),\bF^{\mu\nu}(x)].
}

We now define a perturbative solution $\bA$ that takes values in $\Lie$ rather than some given Lie algebra $L$.  \EOMYM\   leads to the following iteration for $\bA_{\leq n}$, with values in $\Lie_{\leq n}$:
\eqnn\EOMYMit
$$\eqalignno{
\bA^\nu_{\leq n+1} &=\bA_1^\nu+ {\rm proj}_{\Lie_{\leq n+1}}\,\Box^{-1} \left([\bA^\mu_{\leq n}(x),\p_\mu \bA^\nu_{\leq n}(x)] + [\bA_{\leq n\mu}(x),\bF_{\leq n}^{\mu\nu}(x)]\right) \, , &\EOMYMit \cr 
\bF^{\mu\nu}_{\leq n+1}&={\rm proj}_{\Lie_{\leq N+1}} \,2\p^{[\mu}\bA^{\nu]}_{\leq n+1}- [\bA^\mu_{\leq n},\bA^\nu_{\leq n}] .
}$$
where ${\rm proj}_{\Lie_{\leq n+1}}$ projects onto the $ \Lie_{\leq n+1}$ part. The recursion is seeded by 
\eqn\seededby{
\bA^\mu_1=\sum_{i\in \Bbb N} e_i^\mu\exp(ik_i\cdot x) \, i
}
where the letter $i$ replaces the usual generator $t^a_i\in \cg$  and $e_i^\mu$ is the polarization vector of the $i$th gluon. Note that $\Box^{-1}$ acts on momentum eigenstates $\exp(ik\cdot x)$ to give  $-1/k^2$.

To obtain the amplitude, define the multi-particle field $\bA_{n-1}^\mu$, which is the degree $n-1$ part of $\bA_{\leq n-1}^\mu$:
\eqn\particlefield{
\bA_{n-1}^\mu:={\rm proj}_{\Lie_{n-1}} \bA_{\leq n-1}^\mu.
}
In terms of $\bA_{n-1}^\mu$, the $\Lie$-valued version of the amplitude is
\eqn\YMnpointamp{
\cA_n= s_{12...n-1}\,e^{-i k_{12...n-1}\cdot x} \, {e_{n}}_\mu\,\bA^\mu_{n-1}  \, ,
}
where $e_n^\mu$ is the polarization of the $n$th gluon. The colour polarizations $t_1, \ldots, t_n\in \cg$ define the map ${\bf t}:\Lie\rightarrow \cg$, and the YM amplitude is then given by $t_n^a{\bf t}(\cA_n)^a$.
The partial tree amplitudes are given by
\eqn\YMpartialamp{
\cA_n(P,n):=(P,\cA_n),
}
for permutations $P$.

The Berends-Giele current for an ordering $P$ is
\eqn\LieA{
J^\mu_P := (P, \bA^\mu ) \exp(-ik_P\cdot x),
}
where the factor of $\exp(-ik_P\cdot x) $ removes  the $x$-dependence in $(P, \bA^\mu_{|P|} )$. $J_P^\mu$ is linear in each $e_i$ with $i\in P$, with coefficients that depend only on the momenta. Taking inner products  \EOMYMit\ with a word $P$ gives the YM Berends-Giele recursion relation:
\eqnn\cAmrec
$$\eqalignno{
J^\mu_P &= {1\over s_{P}}\!\!\!\sum_{XY=P}\!\!\bigl[ J_{X}^\mu (k_X\cdot  J_{Y})
+ J_{X}^\nu \cF_Y^{\mu\nu} - (X \leftrightarrow Y)\bigr],&\cAmrec\cr
\cF^{\mu\nu}_Y &= k_Y^\mu J_Y^\nu - k_Y^\nu J_Y^\mu
- \sum_{RS=Y}\big(J_R^\mu J_S^\nu - J_R^\nu J_S^\mu \big)
}$$
where $J_i^\mu$ for a letter $i$ is equal to the polarization vector $e^\mu_i$ of the $i$-th gluon.
Here again we have used the following deconcatenation identity, 
\eqn\deconagain{
(P, \Gamma)=\sum_{XY=P} (X,\Gamma_1)(Y,\Gamma_2)-(X,\Gamma_2)(Y,\Gamma_1)\, ,
}
for a word $P$, and a Lie monomial $\Gamma = [\Gamma_1,\Gamma_2]$.

Since $\bA_n\in\Lie$, it follows from Ree's theorem that
\eqn\BGJshuffle{
J^m_{R\shuffle S} = 0\,,
}
cf.\ the discussions in  \BGsym\ and  \Gauge. In terms of $J^\mu_P$, the YM partial tree amplitudes are \BGpaper
\eqn\npt{
\cA^{\rm YM}(Pn) = s_{P}J_{P}\cdot J_n\,,
}
which is equivalent to the earlier definition, \YMpartialamp.

\appendix{B}{The main property of the S bracket}
\applab\mainapp

\noindent This appendix proves the following proposition, from \S\BCJsec:

\proclaim Proposition. For $P,Q\in \Lie^*$, the $S$ bracket satisfies
\eqn\bigclaim{
b(\{P,Q\}) = [b(P),b(Q)]\,,
}
i.e., $b$ maps the $S$-bracket to the Lie bracket.

The proof uses the following definitions. The {\it deconcatenation coproduct}  $\delta: W\rightarrow W\otimes W$ is defined on words $P$ by 
\eqn\deconcop{
\delta (P)=\sum_{XY=P} X\otimes Y\, .
}
Write $\delta'(P)$ to be as above but with the sum restricted to non-empty words $X,Y$.   Further, write
\eqn\deconwedge{
\delta_\wedge (P)=\sum_{XY=P} X\otimes Y-Y\otimes X\,,
}
and similarly for $\delta'_\wedge$. Finally, define the $S$-bracket to act on $\Lie^*\otimes \Lie^*$ as
\eqn\SLieLie{
\{X\otimes Y,Q\}=X\otimes \{Y,Q\}\, , \qquad \{P,X\otimes Y\}=\{ P,X\}\otimes Y\, . 
}

\proclaim Lemma. The deconcatenation of the $S$-bracket is
\eqn\deltacurlyAB{
\delta'\{P,Q\}\sim \{\delta_\wedge P,Q\}+\{P,\delta_\wedge Q\}+ s_{P,Q} P\otimes Q\, .
}
\proof First note that for non-empty $X,Y$
\eqn\deltaXY{
\delta'(XY)=\delta' (X)\cdot ( e\otimes Y) +(X\otimes e)\cdot \delta' (Y),
+X\otimes Y \, ,
}
where $e$ is the empty word.  So
\eqn\deltaPQ{
\delta'\{P,Q\}=\delta'r^*(P)\star \ell^*(Q)+r^*(P)\star\delta' \ell^*(b) + \sum_{P=XiY, Q=ZjW} s_{ij}(X\shuffle \bar Y)i\otimes j(Z\shuffle  W),
}
where we have used the explicit formulas for $\ell^*$ and $r^*$ of \elliP. The KK relations give 
$XiY\sim i(\bar X\shuffle Y)\sim (X\shuffle \bar Y)i$, so the third term in \deltaPQ\ sums to $ s_{P,Q}P\otimes Q$.
The deconcatenation of $\ell^*$ and $r^*$ can be evaluated using \elliP. For example, the deconcatenation of a single term in \elliP is
\eqn\deltaell{
\delta (i \bar X \shuffle Y) = \sum_{X=X_1X_2}\sum_{Y=Y_1Y_2} i \bar X_2\shuffle Y_2 \otimes \bar X_1\shuffle Y_2.
}
Total shuffles vanish in $\Lie^*$. So, in $\Lie^* \otimes \Lie^*$,
\eqn\deltaiXY{
\delta (i \bar X \shuffle Y) = \sum_{Y=Y_1Y_2} i \bar X\shuffle Y_2 \otimes Y_2 +  \sum_{X=X_1X_2} i \bar X_2\shuffle Y \otimes \bar X_1.
}
This can be used to find that
\eqn\deltaellstarP{
\delta' (\ell^*(P))= (\ell^*\otimes 1) \circ \delta'_\wedge(P),
}
and a similar identity for $r^*(Q)$. \qed

\proof (of the proposition)
When $P, Q$ are single letters, \bigclaim\ follows directly. Note that BG recursion can be written as 
\eqn\bPrec{
b(P)=\frac1{s_P} \sum_{X,Y} (X\otimes Y,\delta'(P))\, [b(X),b(Y)]\,,
}
for any homogeneous $P \in \Lie^*$. Substituting $\{P,Q\}$ for $P$ into this recursion, the lemma gives that
\eqn\bPQrec{\eqalign{
s_{PQ} b(\{P,Q\})&=  \sum_{X,Y} \left(X\otimes Y, \{\delta'_\wedge P,Q\}+\{P,\delta'_\wedge (Q)\}+s_{P,Q} P\otimes Q\right) \, [b(X),b(Y)]\, , \cr
&= s_{P,Q} [b(P),b(Q)] \cr
&\qquad +\sum_{P=XY} [b(X), b(\{Y,Q\})] - (X\leftrightarrow Y)  
\cr
&\qquad + \sum_{Q=XY} [b(\{P,X\}), b(Y)] - (X\leftrightarrow Y)  \, .
}
}
By induction, write $b(\{Y,Q\})=[b(Y),B(Q)]$ in the second last line to find
\eqn\bXbYbQ{
\sum_{P=XY} [b(X), [b(Y),b(Q)] - (X\leftrightarrow Y) =\sum_{P=XY}[[b(X),b(Y)],b(Q)] = s_P[b(P),b(Q)]\, ,
}
using the Jacobi identity followed by BG recursion. The same argument applied to the third line shows that the RHS of \bPQrec\ is $[b(P),b(Q)]$ multiplied by $s_{PQ}$. \qed

\appendix{C}{The KLT matrix formula}
\applab\KLTrecap

This appendix obtains again the standard formula  for the KLT matrix,
\eqn\standardSformula{
S^\ell(1A|1B)=\prod_{i=2}^ns_{i,A_i(AB)},
}
where $A_i(AB)$ is the subset of $\{1,2,3,\ldots ,n\}$ containing all letters that both precede $i$ in $1B$ and follow $i$ in $1A$. We explicitly compute $S^\ell(12...n|1A)$ by expanding $\ell[12...n]$, which is in spirit similar to the argument in \Du.

The main idea is to iteratively apply the main property of the $S$-bracket:
\eqn\itap{
[b(P),k] = b(\{P,k\}),
}
for a letter $k$. Using the formula for $\{P,k\}$ (equation \ordinaryfundbcj), this gives
\eqn\itell{
\eqalign{
[\ldots [[1,2]3]\ldots ,n]&= s_{12}[\ldots [b(12),3]\ldots ,n]\cr
&= s_{12}s_{1,3} [\ldots [b(132),4]\ldots ,n]\cr
& \qquad + s_{12} s_{12,3}[\ldots [b(123),4]\ldots ,n]\cr
&=\ldots
}}
The RHS is the nested sum,
\eqn\nestedsum{
s_{12}\left(\sum_{12=A_2B_2} s_{2,A_2} \left(\ldots \left(\sum_{A_{n-1}nB_{n-1}-A_nB_n} s_{n,A_n} b(A_nnB_n)\right)\ldots \right)\right).
}
Here all the words $A_i$ begin with the letter $1$ and by convention $s_{i,A}=0$ when $A$ is empty.  $S^\ell(12\ldots n| 1A)$ is the coefficient of $b(1A)$ in \nestedsum. Reversing the order of the summations gives this coefficient as
\eqn\finalstandard{
S^\ell(12\ldots n| 1A)= \prod_{i=2}^n \left( \sum_{C_i=A_iiB_i} s_{i,A_i}\right)
}
where $C_i$ is the word $1A$ with the letters $1,\ldots ,i-1$ removed.  In other words, $A_i$ is a word in the letters $j$ such that $j<i$ in the ordering $1A$ and $i<j$ in the ordering $12\ldots n$.

\ninerm
\listrefs

\bye

%% file: harvmacMv2.tex


\input amssym.tex 

\def\unredoffs{}
\tolerance=1000\hfuzz=2pt
\catcode`\@=11 
\ifx\hyperdef\UNd@FiNeD\def\hyperdef#1#2#3#4{#4}\def\hyperref#1#2#3#4{#4}\def\href#1#2{#2}\fi
\magnification=1200\unredoffs\baselineskip=16pt plus 2pt minus 1pt
\def\Date#1{\vfill\leftline{#1}\tenpoint\supereject%
\footline={\hss\tenrm\hyperdef\hypernoname{page}\folio\folio\hss}}%

{\count255=\time\divide\count255 by 60 \xdef\hourmin{\number\count255}
 \multiply\count255 by-60\advance\count255 by\time
 \xdef\hourmin{\hourmin:\ifnum\count255<10 0\fi\the\count255}
}
\def\date{\number\day.\number\month.\number\year\ at \hourmin}


\def\nolabels{\def\wrlabeL##1{}\def\eqlabeL##1{}\def\reflabeL##1{}}
\def\writelabels{\def\wrlabeL##1{\leavevmode\vadjust{\rlap{\smash%
{\line{{\escapechar=` \hfill\rlap{\sevenrm\hskip.03in\string##1}}}}}}}%
\def\eqlabeL##1{{\escapechar-1\rlap{\sevenrm\hskip.05in\string##1}}}%
\def\reflabeL##1{\noexpand\llap{\noexpand\sevenrm\string\string\string##1}}}
\nolabels

\global\newcount\secno \global\secno=0
\global\newcount\meqno \global\meqno=1
\def\s@csym{}

\def\newsec#1\par{\global\advance\secno by1%
{\toks0{#1}\message{(\the\secno. \the\toks0)}}%
\global\subsecno=0\eqnres@t\let\s@csym\secsym\xdef\secn@m{\the\secno}\noindent
{\bf\hyperdef\hypernoname{section}{\the\secno}{\the\secno.} #1}%
\writetoca{{\string\hyperref{}{section}{\the\secno}{\bf \the\secno\quad}} {\bf #1}}\par%
\nobreak\medskip\nobreak\noindent\ignorespaces}
\def\eqnres@t{\xdef\secsym{\the\secno.}\global\meqno=1\bigbreak\bigskip}
\def\sequentialequations{\def\eqnres@t{\bigbreak}}\xdef\secsym{}

\global\newcount\subsecno \global\subsecno=0
\def\subsec#1\par{\global\advance\subsecno by1%
{\toks0{#1}\message{(\s@csym\the\subsecno. \the\toks0)}}%
\global\subsubsecno=0%
\ifnum\lastpenalty>9000\else\bigbreak\fi
\noindent{\it\hyperdef\hypernoname{subsection}{\secn@m.\the\subsecno}%
{\secn@m.\the\subsecno.} #1}\writetoca{\string\hskip1.45cm
{\string\hyperref{}{subsection}{\secn@m.\the\subsecno}{\secn@m.\the\subsecno.}}
{#1}}\par\nobreak\medskip\nobreak\noindent\ignorespaces}

\global\newcount\subsubsecno \global\subsubsecno=0
\def\subsubsec#1\par{\global\advance\subsubsecno by1%
{\toks0{#1}\message{(\secn@m.\the\subsecno.\the\subsubsecno. \the\toks0)}}%
\global\subsubsubsecno=0%
\ifnum\lastpenalty>9000\else\bigbreak\fi
\noindent{\it\hyperdef\hypernoname{subsubsection}{\secn@m.\the\subsecno\the\subsubsecno}%
{\secn@m.\the\subsecno.\the\subsubsecno.} #1}
\par\nobreak\medskip\nobreak\noindent\ignorespaces}

\global\newcount\subsubsubsecno \global\subsubsubsecno=0
\def\subsubsubsec#1\par{\global\advance\subsubsubsecno by1%
{\toks0{#1}\message{(\secn@m.\the\subsecno.\the\subsubsecno.\the\subsubsubsecno \the\toks0)}}%
\ifnum\lastpenalty>9000\else\bigbreak\fi
\noindent{\it\hyperdef\hypernoname{subsubsection}{\secn@m.\the\subsecno\the\subsubsecno\the\subsubsubsecno}%
{\secn@m.\the\subsecno.\the\subsubsecno.\the\subsubsubsecno.} #1}%
\par\nobreak\medskip\nobreak\noindent\ignorespaces}


\def\newnewsec#1#2\par{\global\advance\secno by1%
{\toks0{#2}\message{(\secn@m. \the\toks0)}}%
\global\subsecno=0\global\subsubsecno=0\eqnres@t\let\s@csym\secsym\xdef\secn@m{\the\secno}\noindent
\ifnum\lastpenalty>9000\else\bigbreak\fi
\noindent{\bf\hyperdef\hypernoname{section}{\secn@m}{\secn@m.} #2}%
\writetoca{{\string\hyperref{}{section}{\the\secno}{\bf \the\secno\quad}} {\bf #2}}
\DefWarn#1%
\xdef#1{\noexpand\hyperref{}{section}{\the\secno}%
{\the\secno}}\writedef{#1\leftbracket#1}\wrlabeL{#1=#1}%
\par\nobreak\medskip\nobreak\noindent\ignorespaces}

\def\newsubsec#1#2\par{\global\advance\subsecno by1%
{\toks0{#2}\message{(\secn@m.\the\subsecno. \the\toks0)}}%
\global\subsubsecno=0%
\ifnum\lastpenalty>9000\else\bigbreak\fi
\noindent{\it\hyperdef\hypernoname{subsection}{\secn@m.\the\subsecno}%
{\secn@m.\the\subsecno.} #2}
\DefWarn#1%
\xdef#1{\noexpand\hyperref{}{subsection}{\secn@m.\the\subsecno}%
{\secn@m.\the\subsecno}}\writedef{#1\leftbracket#1}\wrlabeL{#1=#1}%
\writetoca{\string\hskip1.45cm
{\string\hyperref{}{subsection}{\secn@m.\the\subsecno}{\secn@m.\the\subsecno.}}
{#2}}%
\par\nobreak\medskip\nobreak\noindent\ignorespaces}

\def\newsubsubsec#1#2\par{\global\advance\subsubsecno by1%
{\toks0{#2}\message{(\secn@m.\the\subsecno.\the\subsubsecno. \the\toks0)}}%
\global\subsubsubsecno=0%
\ifnum\lastpenalty>9000\else\bigbreak\fi
\noindent{\it\hyperdef\hypernoname{subsubsection}{\secn@m.\the\subsecno\the\subsubsecno}%
{\secn@m.\the\subsecno.\the\subsubsecno.} #2}
\DefWarn#1%
\xdef#1{\noexpand\hyperref{}{subsubsection}{\secn@m.\the\subsecno.\the\subsubsecno}%
{\secn@m.\the\subsecno.\the\subsubsecno}}\writedef{#1\leftbracket#1}\wrlabeL{#1=#1}%
\par\nobreak\medskip\nobreak\noindent\ignorespaces}

\def\newsubsubsubsec#1#2\par{\global\advance\subsubsubsecno by1%
{\toks0{#2}\message{(\secn@m.\the\subsecno.\the\subsubsecno.\the\subsubsubsecno \the\toks0)}}%
\ifnum\lastpenalty>9000\else\bigbreak\fi
\noindent{\it\hyperdef\hypernoname{subsubsection}{\secn@m.\the\subsecno\the\subsubsecno\the\subsubsubsecno}%
{\secn@m.\the\subsecno.\the\subsubsecno.\the\subsubsubsecno.} #2}
\DefWarn#1%
\xdef#1{\noexpand\hyperref{}{subsubsubsection}{\secn@m.\the\subsecno.\the\subsubsecno.\the\subsubsubsecno}%
{\secn@m.\the\subsecno.\the\subsubsecno.\the\subsubsubsecno}}\writedef{#1\leftbracket#1}\wrlabeL{#1=#1}%
\par\nobreak\medskip\nobreak\noindent\ignorespaces}

\def\appendix#1#2{\global\meqno=1\global\subsecno=0\global\subsubsecno=0\xdef\secsym{\hbox{#1.}}%
\bigbreak\bigskip\noindent{\bf Appendix \hyperdef\hypernoname{appendix}{#1}%
{#1.} #2}{\toks0{(#1. #2)}\message{\the\toks0}}%
\xdef\s@csym{#1.}\xdef\secn@m{#1}%
\writetoca{{\string\hyperref{}{appendix}{#1}{\bf {#1}\quad}} {\bf #2}}%
\par\nobreak\medskip\nobreak}

%
\def\checkm@de#1#2{\ifmmode{\def\f@rst##1{##1}\hyperdef\hypernoname{equation}%
{#1}{#2}}\else\hyperref{}{equation}{#1}{#2}\fi}
\def\eqnn#1{\DefWarn#1\xdef #1{(\noexpand\relax\noexpand\checkm@de%
{\s@csym\the\meqno}{\secsym\the\meqno})}%
\wrlabeL#1\writedef{#1\leftbracket#1}\global\advance\meqno by1}
\def\f@rst#1{\c@t#1a\em@ark}\def\c@t#1#2\em@ark{#1}
\def\eqna#1{\DefWarn#1\wrlabeL{#1$\{\}$}%
\xdef #1##1{(\noexpand\relax\noexpand\checkm@de%
{\s@csym\the\meqno\noexpand\f@rst{##1}1}{\hbox{$\secsym\the\meqno##1$}})}
\writedef{#1\numbersign1\leftbracket#1{\numbersign1}}\global\advance\meqno by1}
\def\eqn#1#2{\DefWarn#1%
\xdef #1{(\noexpand\hyperref{}{equation}{\s@csym\the\meqno}%
{\secsym\the\meqno})}$$#2\eqno(\hyperdef\hypernoname{equation}%
{\s@csym\the\meqno}{\secsym\the\meqno})\eqlabeL#1$$%
\writedef{#1\leftbracket#1}\global\advance\meqno by1}
\def\xeqn{\expandafter\xe@n}\def\xe@n(#1){#1}
\def\xeqna#1{\expandafter\xe@n#1}
\def\eqns#1{(\e@ns #1{\hbox{}})}
\def\e@ns#1{\ifx\UNd@FiNeD#1\message{eqnlabel \string#1 is undefined.}%
\xdef#1{(?.?)}\fi{\let\hyperref=\relax\xdef\next{#1}}%
\ifx\next\em@rk\def\next{}\else%
\ifx\next#1\xeqn#1\else\def\n@xt{#1}\ifx\n@xt\next#1\else\xeqna#1\fi
\fi\let\next=\e@ns\fi\next}
\def\DefWarn#1{}
%
\newskip\footskip\footskip14pt plus 1pt minus 1pt 
\def\footnotefont{\ninepoint}\def\f@t#1{\footnotefont #1\@foot}
\def\f@@t{\baselineskip\footskip\bgroup\footnotefont\aftergroup\@foot\let\next}
\setbox\strutbox=\hbox{\vrule height9.5pt depth4.5pt width0pt}
\global\newcount\ftno \global\ftno=0
\def\foot{\global\advance\ftno by1\def\foot@rg{\hyperref{}{footnote}%
{\the\ftno}{\the\ftno}\xdef\foot@rg{\noexpand\hyperdef\noexpand\hypernoname%
{footnote}{\the\ftno}{\the\ftno}}}\footnote{$^{\foot@rg}$}}
%
%
%
\global\newcount\refno \global\refno=1
\newwrite\rfile
\def\ref{[\hyperref{}{reference}{\the\refno}{\the\refno}]\nref}
\def\nref#1{\DefWarn#1%
\xdef#1{[\noexpand\hyperref{}{reference}{\the\refno}{\the\refno}]}%
\writedef{#1\leftbracket#1}%
\ifnum\refno=1\immediate\openout\rfile=\jobname.refs\fi
\chardef\wfile=\rfile\immediate\write\rfile{\noexpand\item{[\noexpand\hyperdef%
\noexpand\hypernoname{reference}{\the\refno}{\the\refno}]\ }%
\reflabeL{#1\hskip.31in}\pctsign}\global\advance\refno by1\findarg}
\def\findarg#1#{\begingroup\obeylines\newlinechar=`\^^M\pass@rg}
{\obeylines\gdef\pass@rg#1{\writ@line\relax #1^^M\hbox{}^^M}%
\gdef\writ@line#1^^M{\expandafter\toks0\expandafter{\striprel@x #1}%
\edef\next{\the\toks0}\ifx\next\em@rk\let\next=\endgroup\else\ifx\next\empty%
\else\immediate\write\wfile{\the\toks0}\fi\let\next=\writ@line\fi\next\relax}}
\def\striprel@x#1{} \def\em@rk{\hbox{}}
\def\lref{\begingroup\obeylines\lr@f}
\def\lr@f#1#2{\DefWarn#1\gdef#1{\let#1=\UNd@FiNeD\ref#1{#2}}\endgroup\unskip}
\def\semi{;\hfil\break}
\def\addref#1{\immediate\write\rfile{\noexpand\item{}#1}} 
\def\listrefs{\vfill\supereject\immediate\closeout\rfile\writestoppt
\baselineskip=\footskip\centerline{{\bf References}}\bigskip{\parindent=20pt%
\frenchspacing\escapechar=` \input \jobname.refs\vfill\eject}\nonfrenchspacing}
\def\startrefs#1{\immediate\openout\rfile=\jobname.refs\refno=#1}
\def\xref{\expandafter\xr@f}\def\xr@f[#1]{#1}
\def\refs#1{\count255=1[\r@fs #1{\hbox{}}]}
\def\r@fs#1{\ifx\UNd@FiNeD#1\message{reflabel \string#1 is undefined.}%
\nref#1{need to supply reference \string#1.}\fi%
\vphantom{\hphantom{#1}}{\let\hyperref=\relax\xdef\next{#1}}%
\ifx\next\em@rk\def\next{}%
\else\ifx\next#1\ifodd\count255\relax\xref#1\count255=0\fi%
\else#1\count255=1\fi\let\next=\r@fs\fi\next}
%

%
\newwrite\ffile\global\newcount\figno \global\figno=1
\def\fig{fig.~\hyperref{}{figure}{\the\figno}{\the\figno}\nfig}
\def\nfig#1{\DefWarn#1%
\xdef#1{fig.~\noexpand\hyperref{}{figure}{\the\figno}{\the\figno}}%
\writedef{#1\leftbracket fig.\noexpand~\xfig#1}%
\ifnum\figno=1\immediate\openout\ffile=\jobname.figs\fi\chardef\wfile=\ffile%
{\let\hyperref=\relax
\immediate\write\ffile{\noexpand\medskip\noexpand\item{Fig.\ %
\noexpand\hyperdef\noexpand\hypernoname{figure}{\the\figno}{\the\figno}. }
\reflabeL{#1\hskip.55in}\pctsign}}\global\advance\figno by1\findarg}
\def\xfig{\expandafter\xf@g}\def\xf@g fig.\penalty\@M\ {}
\def\figs#1{figs.~\f@gs #1{\hbox{}}}
\def\f@gs#1{{\let\hyperref=\relax\xdef\next{#1}}\ifx\next\em@rk\def\next{}\else
\ifx\next#1\xfig #1\else#1\fi\let\next=\f@gs\fi\next}
%
\def\figin{\epsfcheck\figin}\def\figins{\epsfcheck\figins}
\def\epsfcheck{\ifx\epsfbox\UnDeFiNeD
\message{(NO epsf.tex, FIGURES WILL BE IGNORED)}
\gdef\figin##1{\vskip2in}\gdef\figins##1{\hskip.5in}
\else\message{(FIGURES WILL BE INCLUDED)}%
\gdef\figin##1{##1}\gdef\figins##1{##1}\fi}
\def\figinsert{\goodbreak\topinsert}
\def\ifig#1#2#3{\DefWarn#1\xdef#1{fig.~\the\figno}
\writedef{#1\leftbracket fig.\noexpand~\the\figno}%
\figinsert\figin{\centerline{#3}}
\smallskip
\leftskip=0pt \rightskip=0pt
\baselineskip12pt\noindent
{{\bf Fig.~\the\figno}\ \ninepoint #2}
\medskip
\global\advance\figno by1\par\endinsert}
\newwrite\lfile
{\escapechar-1\xdef\pctsign{\string\%}\xdef\leftbracket{\string\{}
\xdef\rightbracket{\string\}}\xdef\numbersign{\string\#}}
\def\writedefs{\immediate\openout\lfile=label.defs \def\writedef##1{%
{\let\hyperref=\relax\let\hyperdef=\relax\let\hypernoname=\relax
 \immediate\write\lfile{\string\checkdef\string##1\rightbracket}}}}%
\def\writestop{\def\writestoppt{\immediate\write\lfile{\string\pageno
 \the\pageno\string\startrefs\leftbracket\the\refno\rightbracket
 \string\def\string\secsym\leftbracket\secsym\rightbracket
 \string\secno\the\secno\string\meqno\the\meqno}\immediate\closeout\lfile}}
\def\writestoppt{}\def\writedef#1{}

\def\seclab#1\par{\DefWarn#1%
\xdef #1{\noexpand\hyperref{}{section}{\the\secno}{\the\secno}}%
\writedef{#1\leftbracket#1}\wrlabeL{#1=#1}\par%
\nobreak\medskip\nobreak\noindent\ignorespaces}
\def\subseclab#1\par{\DefWarn#1%
\xdef #1{\noexpand\hyperref{}{subsection}{\the\secno.\the\subsecno}%
{\the\secno.\the\subsecno}}\writedef{#1\leftbracket#1}\wrlabeL{#1=#1}\par%
\nobreak\medskip\nobreak\noindent\ignorespaces}
\def\subsubseclab#1\par{\DefWarn#1%
\xdef#1{\noexpand\hyperref{}{subsubsection}{\the\secno.\the\subsecno.\the\subsubsecno}%
{\the\secno.\the\subsecno.\the\subsubsecno}}\writedef{#1\leftbracket#1}\wrlabeL{#1=#1}\par%
\nobreak\medskip\nobreak\noindent\ignorespaces}
\def\applab#1\par{\DefWarn#1%
\xdef#1{\noexpand\hyperref{}{appendix}{\secn@m}{\secn@m}}%
\writedef{#1\leftbracket#1}\wrlabeL{#1=#1}%
\par\nobreak\medskip\nobreak\noindent\ignorespaces}
\def\appsublab#1{\DefWarn#1%
\xdef #1{\noexpand\hyperref{}{appendix}{\secn@m.\the\subsecno}{\secn@m.\the\subsecno}}%
\writedef{#1\leftbracket#1}\wrlabeL{#1=#1}}
\newwrite\tfile \def\writetoca#1{}
\def\leaderfill{\leaders\hbox to 1em{\hss.\hss}\hfill}
\def\writetoc{\immediate\openout\tfile=\jobname.toc
   \def\writetoca##1{{\edef\next{\write\tfile{\noindent ##1
   \string\leaderfill{
   \string\hyperref{}{page}{\noexpand\number\pageno}%
   {\noexpand\number\pageno}} \par}}\next}}
}
\newread\ch@ckfile
\def\listtoc{\immediate\closeout\tfile\immediate\openin\ch@ckfile=\jobname.toc
\ifeof\ch@ckfile\message{no file \jobname.toc, no table of contents this pass}%
\else\closein\ch@ckfile\centerline{\bf Contents}\nobreak\medskip%
{\baselineskip=15.5pt\footnotefont\parskip=0pt\catcode`\@=11\input\jobname.toc
\catcode`\@=12\bigbreak\bigskip}\fi}
\catcode`\@=12 
\def\tenpoint{\def\rm{\fam0\tenrm}
\textfont0=\tenrm \scriptfont0=\sevenrm \scriptscriptfont0=\fiverm
\textfont1=\teni  \scriptfont1=\seveni  \scriptscriptfont1=\fivei
\textfont2=\tensy \scriptfont2=\sevensy \scriptscriptfont2=\fivesy
\textfont\itfam=\tenit \def\it{\fam\itfam\tenit}\def\footnotefont{\ninepoint}%
\textfont\bffam=\tenbf \def\bf{\fam\bffam\tenbf}\def\sl{\fam\slfam\tensl}\rm}
\font\ninerm=cmr9 \font\sixrm=cmr6 \font\ninei=cmmi9 \font\sixi=cmmi6
\font\ninesy=cmsy9 \font\sixsy=cmsy6 \font\ninebf=cmbx9
\font\nineit=cmti9 \font\ninesl=cmsl9 \skewchar\ninei='177
\skewchar\sixi='177 \skewchar\ninesy='60 \skewchar\sixsy='60
\def\ninepoint{\def\rm{\fam0\ninerm}
\textfont0=\ninerm \scriptfont0=\sixrm \scriptscriptfont0=\fiverm
\textfont1=\ninei \scriptfont1=\sixi \scriptscriptfont1=\fivei
\textfont2=\ninesy \scriptfont2=\sixsy \scriptscriptfont2=\fivesy
\textfont\itfam=\ninei \def\it{\fam\itfam\nineit}\def\sl{\fam\slfam\ninesl}%
\textfont\bffam=\ninebf \def\bf{\fam\bffam\ninebf}\rm}
%
\hyphenation{anom-aly anom-alies coun-ter-term coun-ter-terms}

\def\tikzcaption#1#2{\DefWarn#1\xdef#1{Fig.~\the\figno}
\writedef{#1\leftbracket Fig.\noexpand~\the\figno}%
{
\smallskip
\leftskip=20pt \rightskip=20pt \baselineskip12pt\noindent
{{\bf Fig.~\the\figno}\ \ninepoint #2}
\bigskip
\global\advance\figno by1 \par}}

\def\ntoalpha#1{%
\ifcase#1%
@%
\or A\or B\or C\or D\or E\or F\or G\or H\or I\or J\or K\or L\or M%
\fi
}

\global\newcount\appno \global\appno=1
\def\applab#1{\xdef #1{\ntoalpha{\appno}}\writedef{#1\leftbracket#1}\wrlabeL{#1=#1}
\global\advance\appno by1}

\def\preprint#1 #2\par{\rightline{\vbox{\baselineskip12pt\hbox{#1}\hbox{#2}}}\vskip2cm}
%
\def\title#1\par{\centerline{\bf #1}\nopagenumbers\pageno=0}
\def\author#1\par{\bigskip\bigskip\centerline{#1}}

\newcount\addressno

\def\email#1#2{
\footnote{\null}{\kern-\parindent \llap{$^#1$\hskip1pt}email: #2}}

\def\startcenter{%
  \par
  \begingroup
  \leftskip=0pt plus 1fil
  \rightskip=\leftskip
  \parindent=0pt
  \parfillskip=0pt
}
\def\stopcenter{\endgroup}

\def\address{\bigskip%
  \ifnum\the\addressno=0\else\stopcenter\endgroup\fi
  \advance\addressno by 1%
  \begingroup
  \startcenter
  \it
  \obeylines
  \addressAux
}
\def\addressAux#1{#1}

\def\abstract{\stopcenter\endgroup\bigskip\bigskip\noindent}

\def\Dsl{\,\raise.15ex\hbox{/}\mkern-13.5mu D} 
\def\dsl{\raise.15ex\hbox{/}\kern-.57em\partial}
\def\tr{{\rm tr}} 
\def\boxeqn#1{\vcenter{\vbox{\hrule\hbox{\vrule\kern3pt\vbox{\kern3pt
	\hbox{${\displaystyle #1}$}\kern3pt}\kern3pt\vrule}\hrule}}}


\def\ap{{\alpha^{\prime}}}

\def\a{\alpha}

\def\t{{\theta}}

\def\half{{1\over 2}}
\def\p{{\partial}}

\def\bar{\overline}
\def\({\left(}
\def\){\right)}

\def\cA{{\cal A}}
\def\cF{{\cal F}}

\def\cK{{\cal K}}


\def\bA{{\Bbb A}}

\def\bF{{\Bbb F}}

\def\Box{\square}


\def\len#1{{%
\def\Dlen{\left|\mkern-1mu #1\mkern -0.5mu\right|}%
\def\Sslen{\left|\mkern-1.3mu #1\mkern -1.3mu\right|}%
\def\SSlen{\left|\mkern-2.8mu #1\mkern-1.3mu\right|}%
\mathchoice{\Dlen}{\Dlen}{\Sslen}{\SSlen}}}

\def\sfrac#1/#2{\kern.1em\raise.5ex\hbox{\the\scriptfont0 #1}%
\kern-.1em/\kern-.15em\lower.25ex\hbox{\the\scriptfont0 #2}}

\font\tenshuffle=shuffle10 \font\sevenshuffle=shuffle7 \font\fiveshuffle=shuffle7 at 5pt
\def\shuffle{{%
\def\Dshuffle{\mathbin{\hbox{\tenshuffle\char'001}}}%
\def\Sshuffle{\mathbin{\hbox{\sevenshuffle\char'001}}}%
\def\SSshuffle{\mathbin{\hbox{\fiveshuffle\char'001}}}%
\mathchoice{\Dshuffle}{\Dshuffle}{\Sshuffle}{\SSshuffle}}}


\def\qed{\hbox{\hskip 3pt
\vbox{\hrule\hbox to 7pt{\vrule height 7pt\hfill\vrule}
\hrule}}\hskip3pt}

\overfullrule=0pt\relax

\frenchspacing

\def\checkdef#1#2{
\ifx\UndeFined#1%
	\def#1{#2}
\else
	\immediate\write16{*** BUG ***: the label \string#1 is already defined ***}
\fi
}
\newread\instream

\openin\instream= label.defs
\ifeof\instream\message{No labels in advance yet. Wait till next pass.}
\else\closein\instream \input label.defs
\fi
\writedefs

\def\arXiv:#1].{\hepthStrip#1 \nil}
\def\hepthStrip#1 #2\nil{\href{http://arxiv.org/abs/#1}{arXiv:#1 #2\unskip}].}

\font\frakfont=eufm10 at 10pt

%% file: scrload.tex

\font\tenscr=rsfs10 
\font\sevenscr=rsfs7 
\font\fivescr=rsfs5 
\skewchar\tenscr='177 \skewchar\sevenscr='177 \skewchar\fivescr='177
\newfam\scrfam \textfont\scrfam=\tenscr \scriptfont\scrfam=\sevenscr
\scriptscriptfont\scrfam=\fivescr
\def\scr{\fam\scrfam}